\documentclass[sigconf]{acmart}%

\usepackage{balance}    %
\usepackage{color}
\usepackage{color, colortbl}

\usepackage{multirow}

\usepackage{verbatim}  %

\usepackage{xspace}

\usepackage{graphicx}
\usepackage{xcolor}

\definecolor{Gray}{gray}{0.97}
\definecolor{MedGray}{gray}{0.9}
\definecolor{greytext}{gray}{0.5}
\definecolor{DarkGreen}{rgb}{0.0, 0.5, 0.0}
\definecolor{PFGreen}{rgb}{0.0, 0.5, 0.0}
\definecolor{lightGreen}{rgb}{0.8, 0.9, 0.8}
\definecolor{CadmiumGreen}{rgb}{0.0, 0.42, 0.24}
\definecolor{DarkKhaki}{rgb}{0.74, 0.72, 0.42}
\definecolor{DarkRed}{rgb}{0.7, 0.2, 0.2}
\definecolor{Purple}{rgb}{0.7,0.0,0.7}
\definecolor{Brown}{rgb}{0.7,0.3,0}
\definecolor{Orange}{rgb}{1, 0.5, 0.1}
\definecolor{niceblue}{rgb}{0.0, 0.2, 0.4}

\graphicspath{
{figures/} %
}

\AtBeginDocument{%
  \providecommand\BibTeX{{%
    \normalfont B\kern-0.5em{\scshape i\kern-0.25em b}\kern-0.8em\TeX}}}

\copyrightyear{2024}
\acmYear{2024}
\setcopyright{rightsretained}
\acmConference[CHI '24]{Proceedings of the CHI Conference on Human Factors in Computing Systems}{May 11--16, 2024}{Honolulu, HI, USA}
\acmBooktitle{Proceedings of the CHI Conference on Human Factors in Computing Systems (CHI '24), May 11--16, 2024, Honolulu, HI, USA}
\acmDOI{10.1145/3613904.3642765}
\acmISBN{979-8-4007-0330-0/24/05}

\usepackage{siunitx}

\begin{document}

\title{SolderlessPCB: Reusing Electronic Components in PCB Prototyping through Detachable 3D Printed Housings}
\author{Zeyu Yan}
\affiliation{%
  \institution{University of Maryland, College Park}
  \city{College Park}
  \state{MD}
  \country{USA}
  \postcode{20895}
} 
\email{zeyuy@umd.edu}
\author{Jiasheng Li}
\affiliation{%
  \institution{University of Maryland, College Park}
  \city{College Park}
  \state{MD}
  \country{USA}
  \postcode{20895}
} 
\email{jsli@umd.edu}
\author{Zining Zhang}
\affiliation{%
  \institution{University of Maryland, College Park}
  \city{College Park}
  \state{MD}
  \country{USA}
  \postcode{20895}
} 
\email{znzhang@umd.edu}
\author{Huaishu Peng}
\affiliation{%
  \institution{University of Maryland, College Park}
  \city{College Park}
  \state{MD}
  \country{USA}
  \postcode{20895}
} 
\email{huaishu@umd.edu}

\renewcommand{\shortauthors}{Yan, et al.}

\begin{abstract}

The iterative prototyping process for printed circuit boards (PCBs) frequently employs surface-mounted device (SMD) components, which are often discarded rather than reused due to the challenges associated with desoldering, leading to unnecessary electronic waste. This paper introduces SolderlessPCB, a collection of techniques for solder-free PCB prototyping, specifically designed to promote the recycling and reuse of electronic components. Central to this approach are custom 3D-printable housings that allow SMD components to be mounted onto PCBs without soldering. We detail the design of SolderlessPCB and the experiments conducted to evaluate its design parameters, electrical performance, and durability. To illustrate the potential for reusing SMD components with SolderlessPCB, we discuss two scenarios: the reuse of components from earlier design iterations and from obsolete prototypes. We also provide examples demonstrating that SolderlessPCB can handle high-current applications and is suitable for high-speed data transmission. The paper concludes by discussing the limitations of our approach and suggesting future directions to overcome these challenges.

\end{abstract}

\ccsdesc[500]{Social and professional topics~Sustainability}
\ccsdesc[500]{Hardware~Printed circuit boards}
\ccsdesc[300]{Human-centered computing~Systems and tools for interaction design}

\keywords{PCB prototyping, Sustainability, Reuse, Electronic Component, Soldering}

\begin{teaserfigure}
  \includegraphics[width=\textwidth]{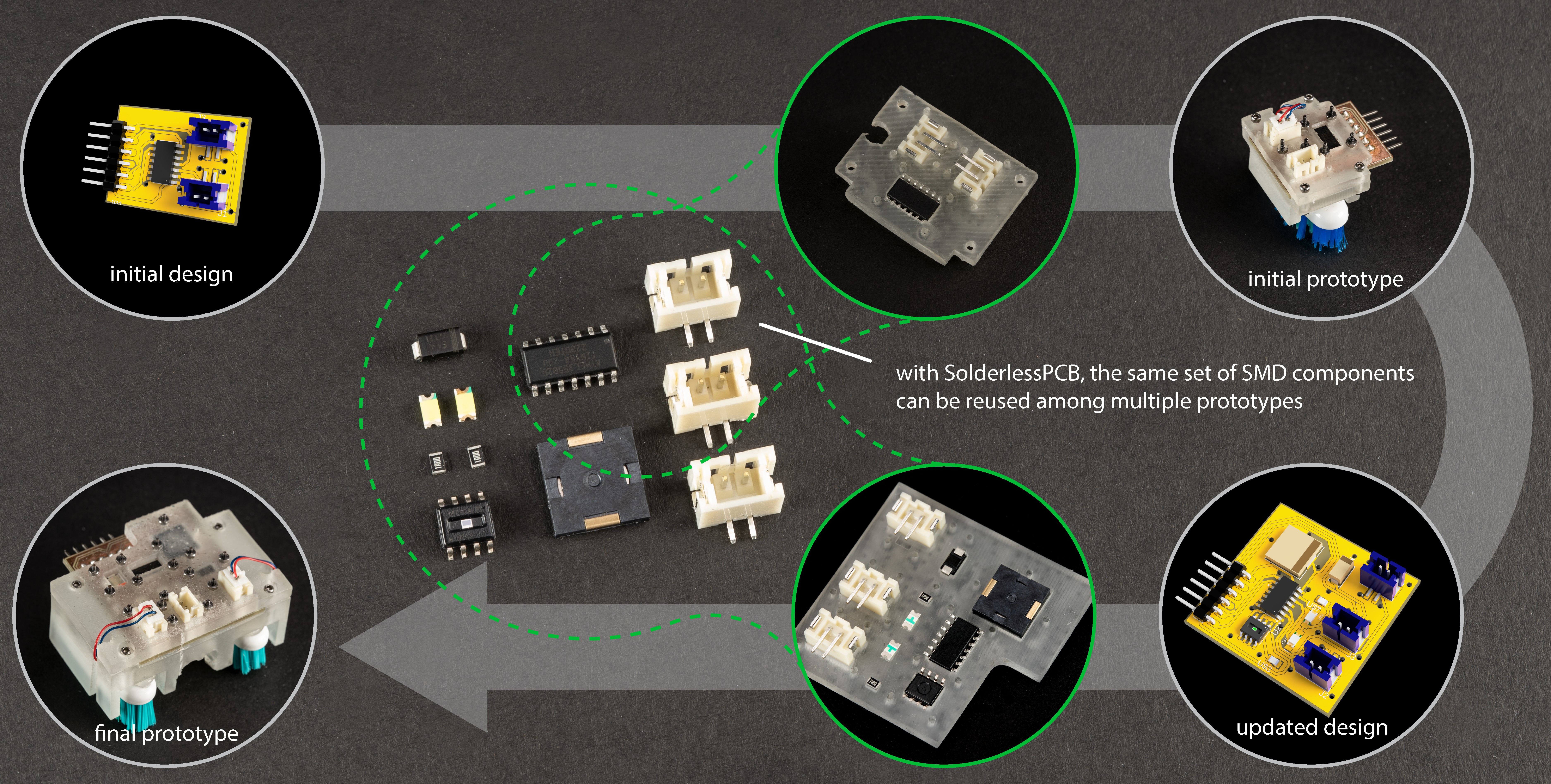}
  \caption{Bristlebot design iterations using SolderlessPCB: the initial design is assembled using a  SolderlessPCB housing; the updated design is assembled into the final prototype by reusing SMD components harvested from the initial prototype.}
  \label{fig:teaser}
  \Description{Figure 1 illustrates the iterative design process for the Bristlebot. Users start with an initial design and assemble it using a solderless PCB housing. Then, they update its design and reuse some SMD (Surface-Mount Device) components from the initial prototype.}
\end{teaserfigure}

\maketitle

\section{Introduction}

Circuit prototyping is a key step in both product design and research. The process often involves multiple rounds of iteration and testing, beginning with circuit validation using breadboards and breakout boards, and later transitioning to custom printed circuit boards (PCBs).

The two stages of circuit prototyping have distinct differences. Breadboard circuitry allows engineers and designers to quickly assemble a circuit by inserting through-hole components and jumper wires into a standardized baseboard. It offers the flexibility to easily modify the circuit by reconnecting different parts or replacing electronic components if mistakes are identified.

In contrast, PCB prototyping is less flexible. Since it usually occurs closer to the final stages of the design process, the form factor of the circuit becomes critical. As a result, the electronic components used in PCB prototyping are often surface-mounted (SMD) with small footprints, requiring assembly through a semi-permanent bonding method—soldering. As such, PCB prototyping boards are often one-off creations. When errors are identified during prototyping, or when a PCB assembly is no longer needed, these PCB boards are often not disassembled like with breadboard prototyping. Instead, new PCBs are produced and assembled with an entirely new suite of electronic components. The components on the old prototyping boards, even though they mostly function perfectly, are not reused or salvaged, leading to unnecessary e-waste. 

This paper explores an alternative approach to traditional PCB prototyping, aimed at promoting the reuse of still-functional SMD components. The key to our idea is a set of 3D-printable housings that can mechanically mount SMD components onto custom PCBs. This eliminates the need for soldering and desoldering during PCB assembly, thereby allowing designers and engineers the flexibility to interchange SMD components similarly to how they would with through-hole components in breadboard prototyping. This approach enables the reuse or replacement of SMD components, while also allowing the exploration of the comparably small form factor of custom PCBs. We call our prototyping technique \textit{SolderlessPCB}.

In the remainder of the paper, we first report on a formative interview exploring how designers and engineers currently conduct circuit prototyping. The design implications drawn from the interview guide our work on SolderlessPCB. We detail the key design considerations and the fabrication method for the 3D-printable housings. We then present a series of experiments on resistance, impedance reproducibility, and high-frequency signal loss in circuits made with SolderlessPCB. These experiments characterize the electrical performance and stability of our approach. To contextualize SolderlessPCB, we present two scenarios in which it can encourage the reuse of electronic components during prototyping, accompanied by three additional examples: a mug heater, a mini game console with an OLED display, and an FTDI code-uploading module, showcasing the applicability of our method. We conclude with a discussion on the limitations of our approach and our thoughts on future research directions.

\section{Related Work}

\subsection{Sustainability in HCI: Making and Prototyping}

Sustainable Interaction Design(SID)~\cite{10.1145/1240624.1240705} highlights the importance of considering environmental impacts and human behaviors in designing interaction technologies, particularly regarding the material outcomes of physical making.
Subsequent works (e.g., ~\cite{10.1145/2851581.2856497, 10.1145/3338103.3338109, 10.1145/3080556.3080568}) have explored the impact of various design paradigms on sustainability.
In particular, prior efforts have demonstrated methods to reduce the environmental impact of waste generated during the design process by utilizing decomposable and recyclable materials~\cite{10.1145/3491102.3502007, 13158115420180801, 10.1145/3563657.3595983, 10.1145/3544549.3583927} or bio-materials~\cite{10.1145/3569009.3572805, 10.1145/3526113.3545623, 10.1145/3569009.3572798, jin2016chitin, 10.1145/3490149.3501308}.
However, materials with unique properties are often not easily replaceable with eco-friendly alternatives, which may require special fabrication processes \cite{8584034} or certain levels of manual manipulations \cite{10.1145/3341163.3346938, 10.1145/3532106.3533494, bell2022reclaym}.

Besides using alternative materials, much research has explored the concepts of repairing and repurposing wasted materials, or ``unmake'' artifacts that are beyond their design lifespan~\cite{10.1145/2556288.2557332, 10.1145/3322276.3322320, 10.1145/3586183.3606745, maestri2011understanding, 10.1145/3411764.3445529, kumar_kumari_sadasivam_goswami_2021, 10.1145/1978942.1979292}. For example, FabricatINK~\cite{10.1145/3491102.3501844} has proposed making personal, bespoke displays using electronic ink from upcycled E-Readers. Unfabricate~\cite{10.1145/3313831.3376227} has demonstrated design practices for reclaiming and reusing materials in the context of smart textiles. 

Recently, HCI researchers have also argued that the design iteration during the prototyping process should also be viewed from a sustainability perspective~\cite{10.1145/3357236.3395510, yan2023future}. 
As most prototyping processes involve multiple iterations, modules and materials are often integrated into the interim results but are rarely reused or recovered~\cite{yan2023future}. In other words, their lifespan is as short as the duration of the testing process for each prototyping iteration. In this research, we examine the prototyping process of PCBs and propose a suite of techniques aimed at reusing the electronic components involved.
We hope this research will spark discussions revolving around reuse and recycling during electronic design iterations.

\subsection{Supporting the Reuse of Electronics}
Research on sustainability in electronics, especially PCBs, extends across multiple domains and encompasses various stages of the PCB lifecycle. 

A common set of industrial approaches for recycling electronics involves extracting raw materials from PCB scrap. Both chemical and mechanical techniques are employed to reclaim valuable materials, including refractory metals and elements from the platinum group, found in standard PCB waste~\cite{C8GC03688H, su131810357}. New recycling methods are also being developed to extract valuable materials from several more specific types of semiconductors, such as ICs and LEDs~\cite{doi:10.1021/acssuschemeng.9b07006, doi:10.1021/ja208577j}. However, these methods have yet to demonstrate generalizability across a wide range of components.

Besides extracting raw materials from discarded PCBs, various research efforts have explored new materials that can make functional PCBs more sustainable. For example, several works have shown that paper~\cite{10.1145/3381013, 10.1145/2493432.2493486, 10.1145/3173574.3174143}, wood~\cite{10.1145/3474349.3480191}, and water-soluble materials~\cite{bharath2020novel, jiva, cheng2023functional, 10.1145/3161165} can all serve as circuit substrates, which can be easily recycled after their lifecycle. Similarly, transient electronics, utilizing novel composite materials, can be engineered to actively or passively degrade into the environment~\cite{https://doi.org/10.1002/adfm.201301847, https://doi.org/10.1002/adma.201403164, vasquez2019plastic, song2023vim}.

From a systems perspective, HCI researchers have explored alternative methods to encourage the reuse of electronic components. For example, CurveBoards~\cite{10.1145/3313831.3376617} has proposed a custom-shaped breadboard design that is versatile for rapid prototyping, enabling the reuse of through-hole components. 
Shorter et al.~\cite{10.1145/2598510.2602965} have demonstrated methods that utilize conductive everyday objects, such as binder clips, to create circuits that can be easily disassembled, albeit with limitations on the resolution and size of the resulting circuits.
ecoEDA ~\cite{10.1145/3586183.3606745} has demonstrated how interactive circuit design software, by integrating early-stage suggestions for utilizing recyclable electronic components from stock PCBs, can facilitate the reuse of electronics throughout the design process. Like ecoEDA, our work also aim to encourage the reuse of electronic components during the PCB prototyping stage. However, our focus is on simplifying the assembly and disassembly of SMD components through the use of a non-soldered mounting method.
 
\subsection{Circuit Assembly and Disassembly}

While recent HCI research has explored the use of CO\textsubscript{2} and fiber lasers to facilitate the soldering and assembly of PCBs~\cite{10.1145/3411764.3445692, 10.1145/3526113.3545652}, the removal and desoldering of electronic components from PCBs remains time-consuming and energy-intensive~\cite{HAO2020104787}.

In the industry, heating methods such as infrared, solder baths, and hot fluids are often used for desoldering~\cite{su131810357}, but they require expensive equipment and are neither environmentally friendly nor safe for health~\cite{NI2014354, WANG201180}. Other works have examined the use of conductive epoxy, rather than solder paste, as a method for circuit assembly~\cite{aradhana2020review}. However, while epoxy does not require the high melting temperatures as with soldered PCBs, it is still difficult to remove without damaging the components.

Our work introduces a new mounting method for PCB assembly, thus entirely eliminates the need of soldering, reduce the risk of damaging SMD components and facilitate solder-free replacement and reuse of these components.

\section{Understanding the Current PCB Prototyping Practice}
To better understand current practices and challenges in PCB prototyping, we conducted semi-structured interviews with experienced PCB designers.

\subsection{Method}
We recruited five participants (four male, one female, aged 26-42) through local makerspaces' email lists and social media platforms. All participants have at least five years of experience in PCB design and prototyping. We visited their labs and makerspaces to conduct in-person, semi-structured interviews. During these interviews, we asked about their general practices of circuit prototyping, as well as their approaches to handling used electronic components. This included both the validation of circuit designs (e.g., using a breadboard) and the process of making PCB prototypes. 
Each interview lasted between 60 and 90 minutes. Participants who successfully completed our interviews were compensated at a rate of \$15 per hour. 
The interviews were audio-recorded and transcribed for analysis. 
Thematic analysis ~\cite{doi:10.1191/1478088706qp063oa} was conducted individually by two investigators and unified through discussion to derive proper findings.
In the rest of this section, we present the key findings from the interviews, which informed the design of SolderlessPCB.

\subsection{Findings}

\subsubsection{Circuit validation with breadboarding}

While all interviewees prototype circuit boards for various objectives, including research, robotic product development, and personal projects, they reported an inevitable circuit validation phase to gain confidence in their circuit diagram.
According to the interviewees, breadboarding is the most commonly used technique to conduct circuit validation, as it allows for easy modifications when necessary.
By the end of this phase, over 60\% of the components are stored for future use. This includes larger components that are placed on the breadboard with DIP packages, such as breakout modules, transducers, and microphones, as well as through-hole components like transistors and LEDs.
One participant stated, \textit{``The nature of my projects is more or less related, so a lot of components will be reused between projects for breadboards.''}
Additionally, apart from small components like resistors and capacitors, which are difficult to have their pins straightened after use, people agree that it is easy and efficient to recycle these components from the breadboard for future use.

\subsubsection{PCB prototyping and assembly}

After the circuit diagram has been verified, PCBs are made and assembled for long-term use.
According to the interviewees, PCBs are either outsourced or made in-house using CNC machines, laser cutters, or chemical etching processes. All of them revealed that more than 80\% of the components used in these PCB assemblies are SMD components.
Two of the interviewees had experience ordering PCB assemblies with components pre-installed by the manufacturer. However, due to differences in cost-effectiveness and lead time between ordering pre-assembled boards and bare PCBs, they ultimately converged their decision with the rest of the interviewees. They chose to make or purchase bare PCB boards and components separately and conduct PCB assembly themselves using soldering irons, heat guns, or reflow ovens.

Despite their experience working with SMD components, none of the interviewees chose to desolder them for reuse between iterations or projects, except in the case of expensive or hard-to-find components.
Interviewees highlighted several challenges in reusing SMD components.
For example, one interviewee mentioned, \textit{``Many components have plastic housing or plastic parts in their packaging. When desoldering them using a heat gun, it is easy to mess them up, and the components end up not being usable anyway.''}
Besides the desoldering techniques, cleaning the leftover solder also presents challenges.
\textit{``Sometimes, it's hard to make sure all pins are properly cleaned after desoldering, so when using components like that, it is hard to solder them again,''} said a circuit designer with eight years of PCB prototyping experience.
Reportedly, less than 15\% of components in PCB assemblies get desoldered and reused.

\subsection{Design Implications}

The interviews confirmed that the low incentive for recycling electronic components during PCB prototyping is largely due to the extensive effort required to desolder components from existing PCBs. We argue that the tedious and unreliable process could be significantly alleviated if the assembly process did not require soldering the components onto the PCB in the first place.

Inspired by the principles of commonly used breadboards~\cite{breadboardExample} and conventional in-circuit testing methods~\cite{buckroyd2015circuit}, we propose assembling SMD components onto PCBs by applying external pressure to the conductive pads. 
Ideally, the agent applying this pressure would eliminate the need for solder, thereby facilitating a smoother and less cumbersome disassembly process, which in turn makes it easier to reuse and recycle SMD components. 
In the following sections, we detail our method of solder-free PCB assembly.

\section{SolderlessPCB} \label{main}

We develop SolderlessPCB, a novel rapid PCB prototyping technique that facilitates easy assembly and disassembly of PCBs as well as the reuse of SMD components.

\begin{figure}[h]
  
  \includegraphics[width=\columnwidth]{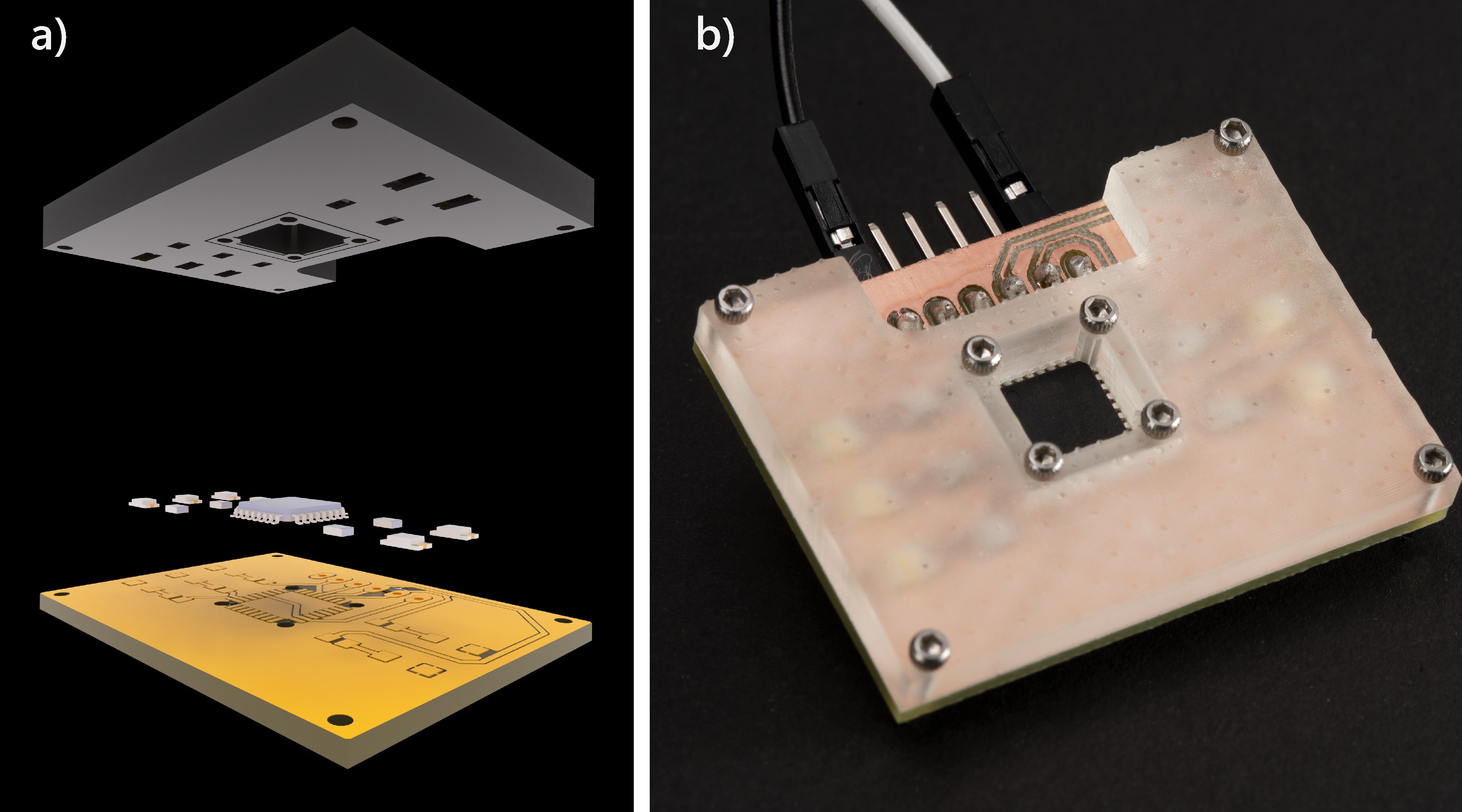}
  \caption{SolderlessPCB: a) rendering of an exploded view of a SolderlessPCB assembly. b) a photo of a SolderlessPCB assembly.}
  \Description{Figure 2: This figure contains two images labeled 'a' and 'b', illustrating the exploded view of SolderlessPCB. Image 'a' shows the exploded view of a SolderlessPCB assembly. The 3D-printable housing on the top part, the SMD components in the middle, and the PCB at the bottom. Image 'b' shows the assembled PCB using SolderlessPCB approach.}
  \label{fig:principle}
\end{figure}

Figure~\ref{fig:principle} illustrates the exploded view of a PCB assembly made with SolderlessPCB.
As shown in the figure, the key to our approach is a custom 3D-printable housing that secures small SMD components to a PCB. 
The housing includes custom cavities sized to fit the SMD components and positioned to align with their locations on the PCB. 
End-users can place all the SMD components inside these cavities and then directly bolt the entire housing to the PCB all at once. 
The housing firmly presses the SMD components against the PCB, forming the electrical connection with its internal tab structures, which also compensate for any height differences among the components. 
With this technique, the PCB assembly process becomes solder-free. In turn, the process for reusing electronic components from assembled PCBs will involve only the unscrewing of the bolts and nuts to release the SMD components from the board.

In the remainder of Section~\ref{main}, we document the detailed designs of SolderlessPCB, which include: 1) the housing anchoring mechanism; 2) the cavity design for various types of electronic components; and 3) the fabrication and design workflow for creating a SolderlessPCB prototype.

\subsection{Housing Anchoring and Pressure Generation}
The key to a successful SolderlessPCB assembly is establishing a reliable electrical connection between the electronic components and the FR-4 PCB baseboard. 
To maintain such a connection, the custom housing needs to exert sufficient pressure to keep the electronic components pressed against the baseboard. 
We explored two sets of housing anchoring mechanisms: snap-fit and screw bolting.  

\subsubsection{Snap-fit}
\begin{figure}[h]
  \includegraphics[width=\columnwidth]{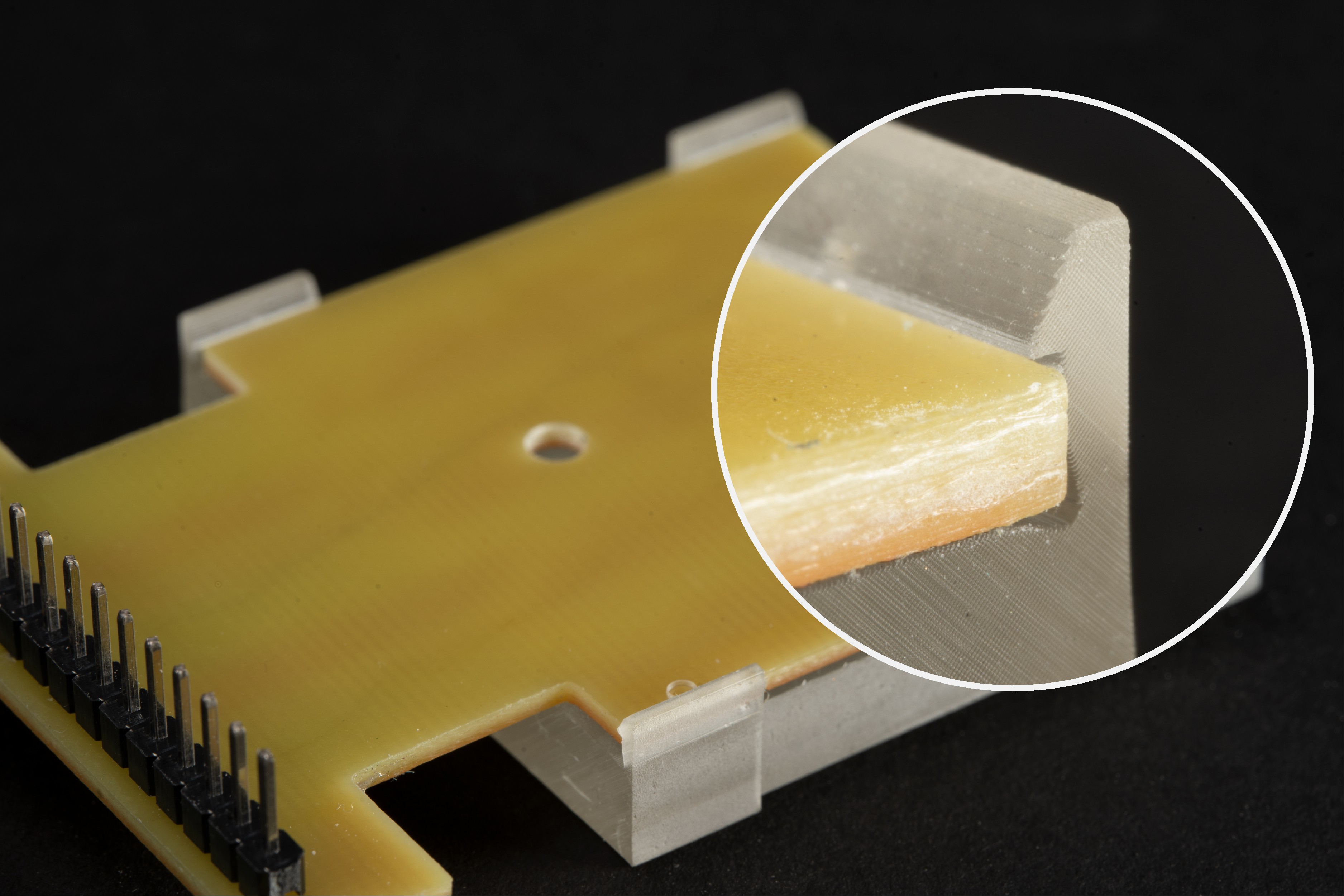}
  \caption{Snap-fit housing PCB assembly with zoomed-in view.}
  \Description{Figure 3: This figure shows the snip-fit design of the housing. The clips achieve with the FR-4 baseboard.}
  \label{fig:clips}
\end{figure}

Snap-fit anchoring consists of four or more custom clips located at the four corners and along the edges of a housing, as shown in Figure~\ref{fig:clips}. 
These clips are designed with groove heights that exactly match the thickness of an FR-4 baseboard. 
As a result, the housing can achieve a tight fit with a baseboard without requiring any additional manual process beyond simply pressing them together. 

While snap-fit is the easiest assembly method, our experiments have shown that this anchoring mechanism does not provide uniform electrical connections across the entire area of a PCB. 
Because the downward force is primarily generated through clipping, components situated closer to the edges receive greater force than those nearer to the center of the baseboard. 
Although allocating more slots for snap-fit mechanisms would improve the overall pressure generated, it's important to note that these slots take up a considerable amount of circuit board area to accommodate the compliant snap-fit tabs. Each slot occupies around \SI{9}{\square\milli\meter} of area.
As a result, snap-fit housing will either severely interfere with PCB design and routing, or not provide reliable electrical connections across the board unless the prototyping board is sufficiently small. 
Due to this major reliability issue, we will not recommend this design overall.

\subsubsection{Screw bolting}\label{bolting}
\begin{figure}[h]
  
  \includegraphics[width=\columnwidth]{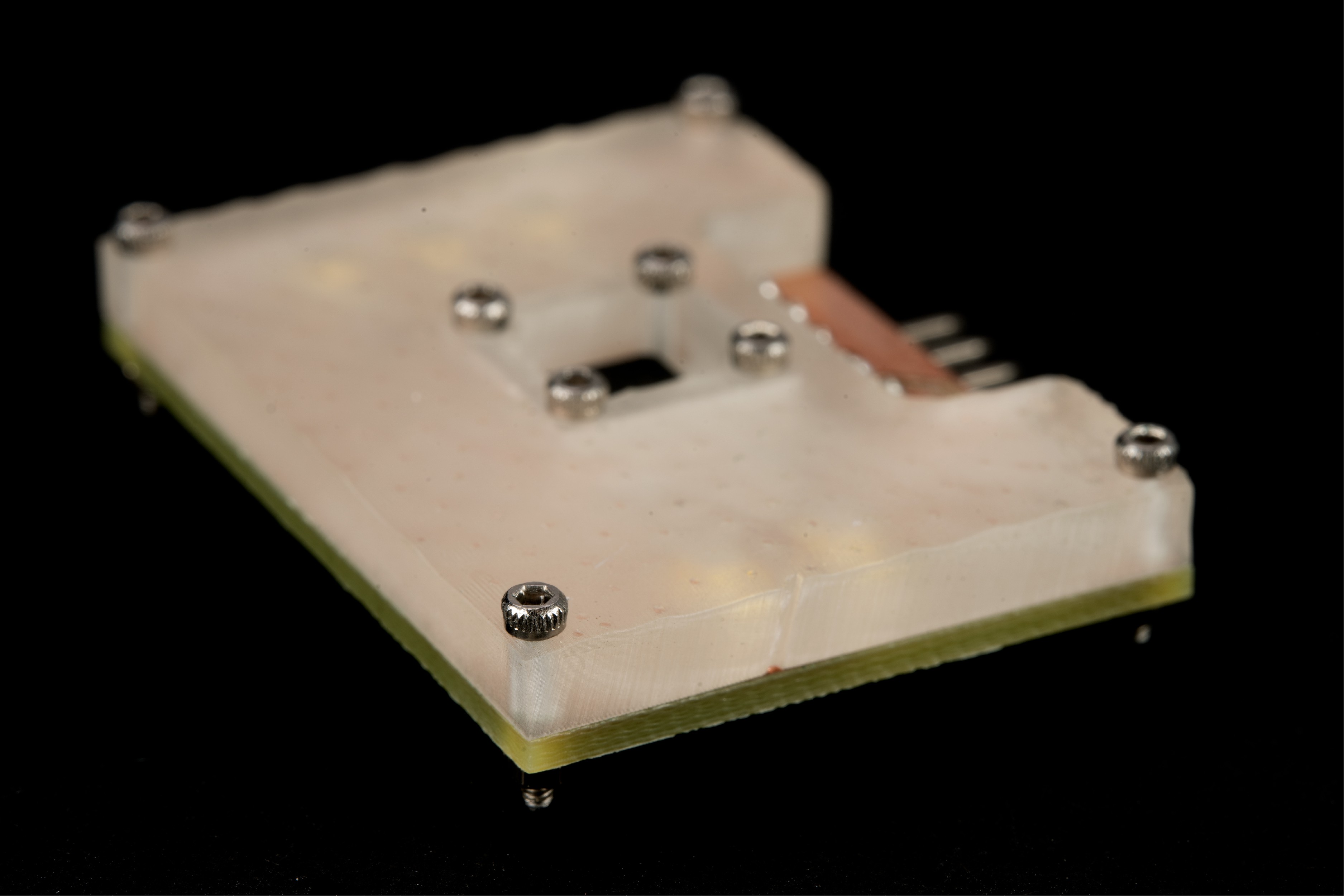}
  \caption{SolderlessPCB assembly using bolts.}
  \Description{Figure 4: This figure shows a SolderlessPCB using bolts.}
  \label{fig:assembly}
\end{figure}

Our second approach uses bolts and nuts with diameters as small as \SI{1}{\milli\meter} to serve as the primary hardware for attaching the housing to the PCB baseboard, as shown in Figure~\ref{fig:assembly}.
While bolting requires additional manual assembly effort, it allows us to distribute fixture hardware, and thus the downward force, evenly across the entire area of a PCB, regardless of their size. 
Since each bolt occupies only \SI{0.8}{\square\milli\meter} of area per hole on the board, its footprint does not significantly interfere with circuit trace routing. 
To ensure optimal pressure generation without damaging the housing, we apply a torque of \SI{0.01}{\newton\meter} to each of the nuts.
Unless otherwise specified, all the examples in the paper are based on the screw bolting mechanism.

\subsection{Cavity Design}

In addition to generating sufficient downward force, the housing must also accommodate a variety of SMD components with different packages and reliably hold them in place mechanically.
Common SMD components include: 1) two-terminal components with regulated package standards, such as resistors, capacitors, and LEDs; 2) integrated circuits (ICs) with standardized packages, such as microcontrollers, IMUs and transistors; and 3) non-standard SMD components like battery holders, terminal connectors, and various specialized sensors. 
We design different types of cavities for each type of component to optimize housing performance. We document these designs in detail below.

\subsubsection{Two-terminal components} \label{0805}

Two-terminal components are among the most commonly seen SMD components. 
They come in a series of numerically ordered package designs, with their dimensions in length and width measured in hundredths of an inch. 
Packages sized 0603, 0805, and 1206 are commonly used for PCB prototyping.

Due to their small size, these two-terminal components often have a degree of height variance. 
While this tolerance does not affect their performance, as they are designed to be soldered onto PCBs, it can pose a challenge for solderless assembly. 
Inconsistent height tolerances can affect the pressure applied during assembly. 
For example, if a component is slightly lower than its expected design height, it may not receive enough pressure to establish a reliable contact with the baseboard. Conversely, if the component is taller than expected, the housing may bend away from the PCB in that area. 
This could potentially allow neighboring components more vertical movement, resulting in loose connections.

\begin{figure}[h]
  
  \includegraphics[width=\columnwidth]{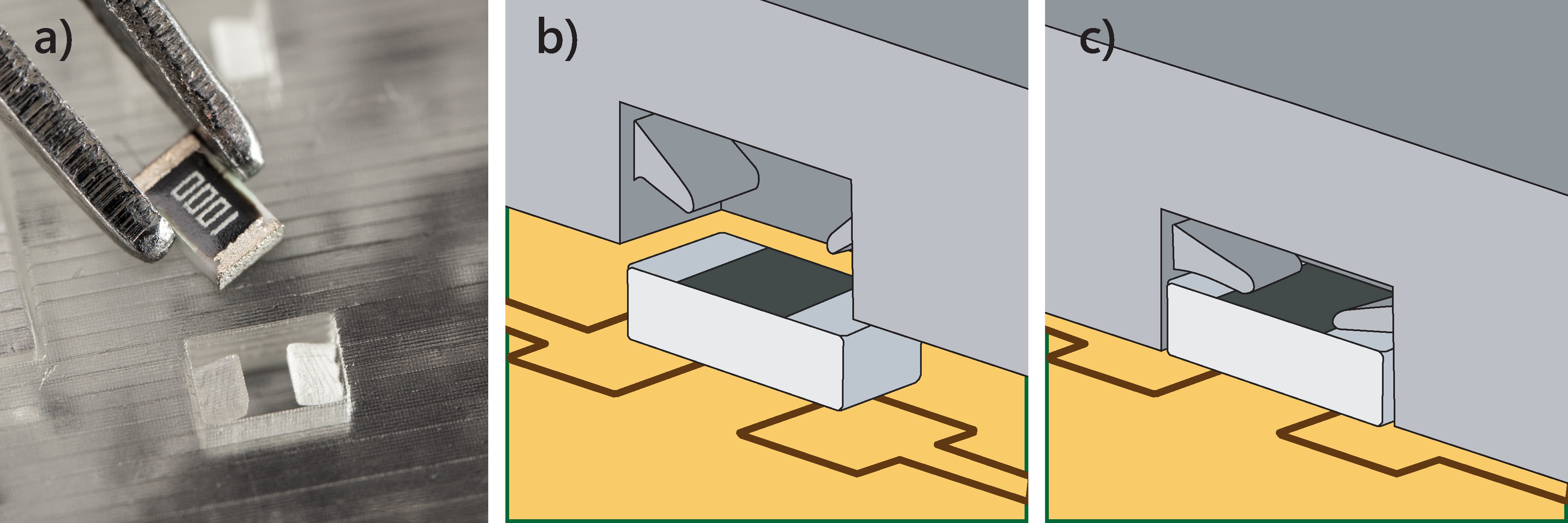}
  \caption{Flexible tab structure accommodating two-terminal components: a) a 0805 component with a housing cavity featuring flexible tabs, b) an exploded view illustration of the housing, a two-terminal component, and the PCB, c) an illustration of the housing pressing the component onto the PCB using flexible tabs.}
  \Description{Figure 5: This figure contains three images labeled in 'a', 'b', and 'c' showing the flexible tab structure accommodating two-terminal components. Image 'a' shows a 0805 component being placed in its housing cavity featuring flexible tabs. Image 'b' shows an exploded view illustration of the housing, a two-terminal component and PCB. Image 'c' shows the housing is pressing the component onto the PCB with flexible tabs.}
  \label{fig:tabs}
\end{figure}

We propose a cubic cavity design with a pair of flexible tabs to compensate for possible height variations caused by manufacturing tolerances (Figure~\ref{fig:tabs}). 
These flexible tabs are located on the side walls of the cavity, angled at \(30^{\circ}\) towards the PCB baseboard.
Upon assembly, the tabs will deform inward, towards the cavity's ceiling. 
Depending on the actual height of each two-terminal component, they will bend to varying degrees (Figure~\ref{fig:tabs}). 
This design will ensure adequate pressure between the components and the PCB, while at the same time absorbing energy through tab deformation to prevent the housing from bending away from the PCB.

\subsubsection{Integrated circuits}

In addition to two-terminal components, the majority of commonly used ICs for rapid prototyping are also available in standardized package designs. 
These include the Small Outline Integrated Circuit (SOIC), Thin Shrink Small Outline Package (TSSOP), Thin Quad Flat Package (TQFP), Ball Grid Array (BGA), and Quad Flat No-Lead Package (QFN).
We categorize these package designs into two main types: Type 1, which includes SOIC, TSSOP, and TQFP, features pins extending from the plastic enclosure; and Type 2, which includes BGA and QFN, where pins are exposed only on the bottom of the chips.
These two types of packages present different challenges for cavity design. We address them separately.

\begin{figure}[h]
  
  \includegraphics[width=\columnwidth]{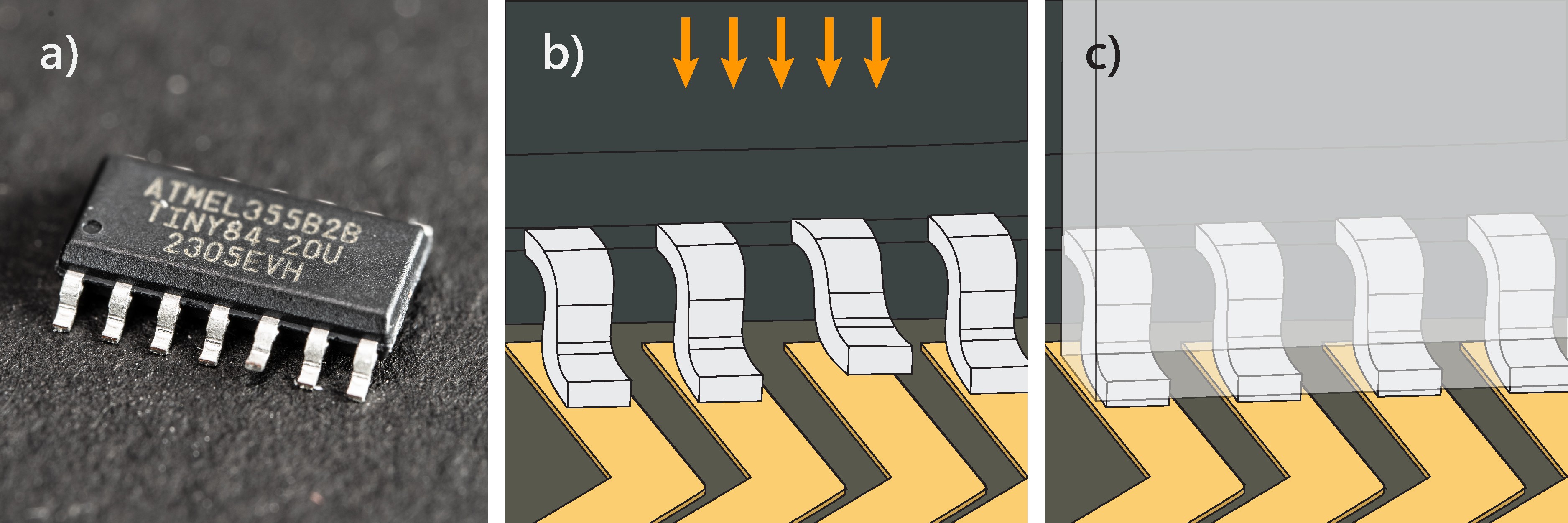}
  \caption{Type 1 IC: a) IC with extended metal pins; b) illustration of unsettled extended pins when pressure is applied at the center; c) illustration of all pins complying when the housing structure is pressed directly upon them.}
  \Description{Figure 6: This figure contains three images labeled in 'a', 'b', and 'c' showing the Type 1 IC. Image 'a' shows a photo of a Type 1 IC. Image 'b' shows an illustration of unsettled extended pins when pressure is applied at the center. Image 'c' shows an illustration of all pins complying when the housing structure is pressed directly upon them.}
  \label{fig:chip1}
\end{figure}

For Type 1 components, the extended metal pins are generally at the same height, although there may be some manufacturing tolerance errors. 
When a downward force is applied solely to the center of the IC enclosure, as with the cavity designed for two-terminal components, the pressure will not be evenly distributed across all the pins of the IC (Figure \ref{fig:chip1}).
Pins that are initially higher than others will not experience the same pressure as the lower ones and may not sufficiently connect to the baseboard.
To account for potential manufacturing discrepancies, our cavity design for Type 1 ICs includes additional solid volumes at the top of each row of metal pins. 
These structures directly press on the metal pins, rather than applying pressure to the center of the IC enclosure (Figure \ref{fig:chip1}c). 
This approach ensures that all metal pins receive downward force, regardless of their individual height differences.

For Type 2 components, an exact negative volume is provided in the housing to accommodate the component, since the exposed metal connectors are not flexible and therefore do not require height compensation.

\begin{figure}[h]
  
  \includegraphics[width=\columnwidth]{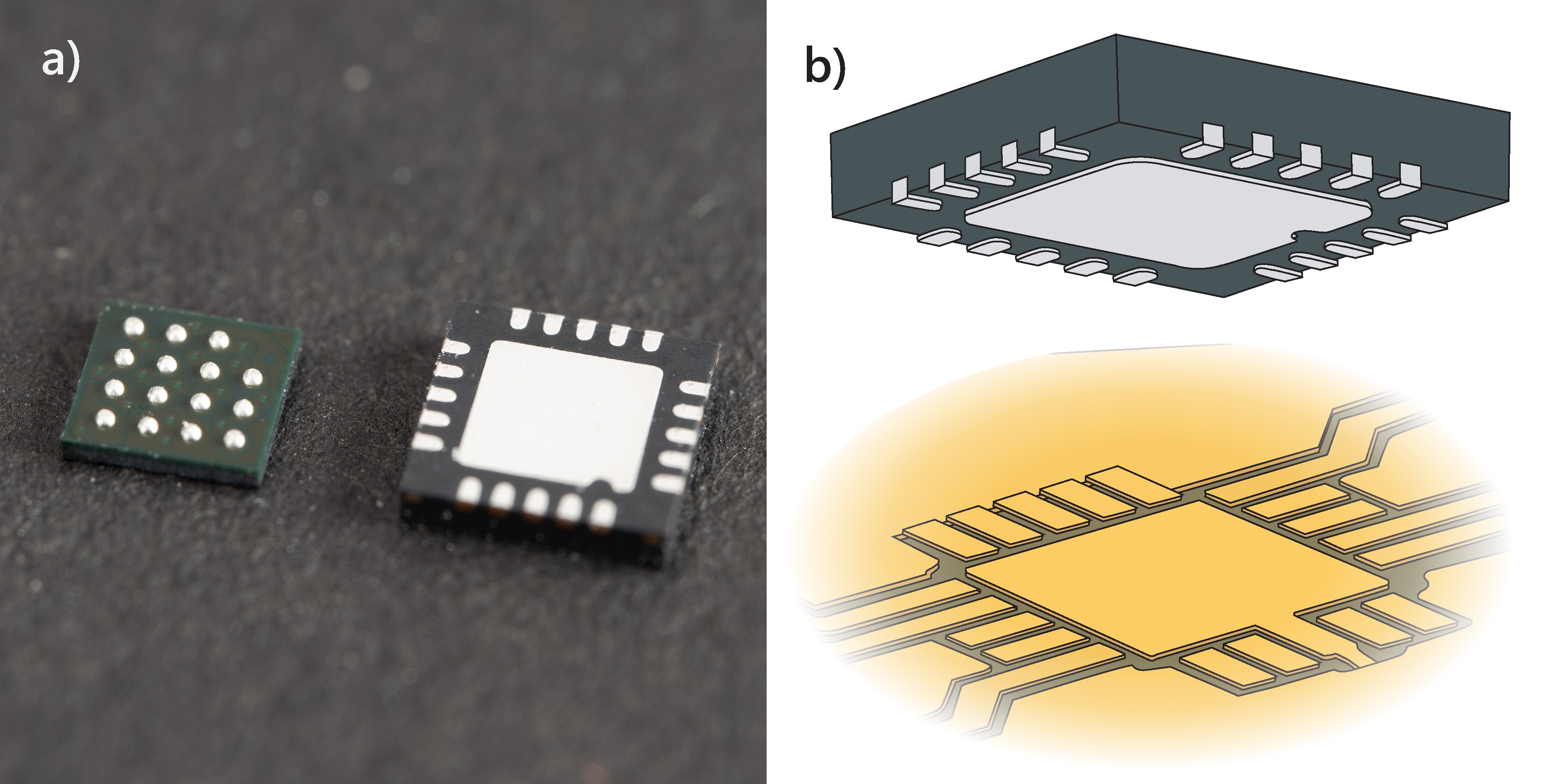}
  \caption{Type 2 ICs: a) IC with hidden pins on the bottom (bottom view); b) illustration of Type 2 ICs working with conductors higher than clearance areas.}
  \Description{Figure 7: This figure contains two images labeled in 'a' and 'b' showing the Type 2 IC. Image 'a' shows a photo of two Type 2 ICs with hidden pins on the bottom. Image 'b' illustrates Type 2 ICs working with conductors higher than clearance areas. }
  \label{fig:chip2}
\end{figure}

However, one challenge for Type 2 components is that the exposed metal connectors are not always the lowest point of the package design. Therefore, when pressing downwards, these metal connectors may not make contact with the baseboard.  
This can be mitigated by deliberately making the copper pads on the PCB baseboard taller than the surrounding areas, as illustrated in Figure \ref{fig:chip2}.

\begin{figure}[h]
  
  \includegraphics[width=\columnwidth]{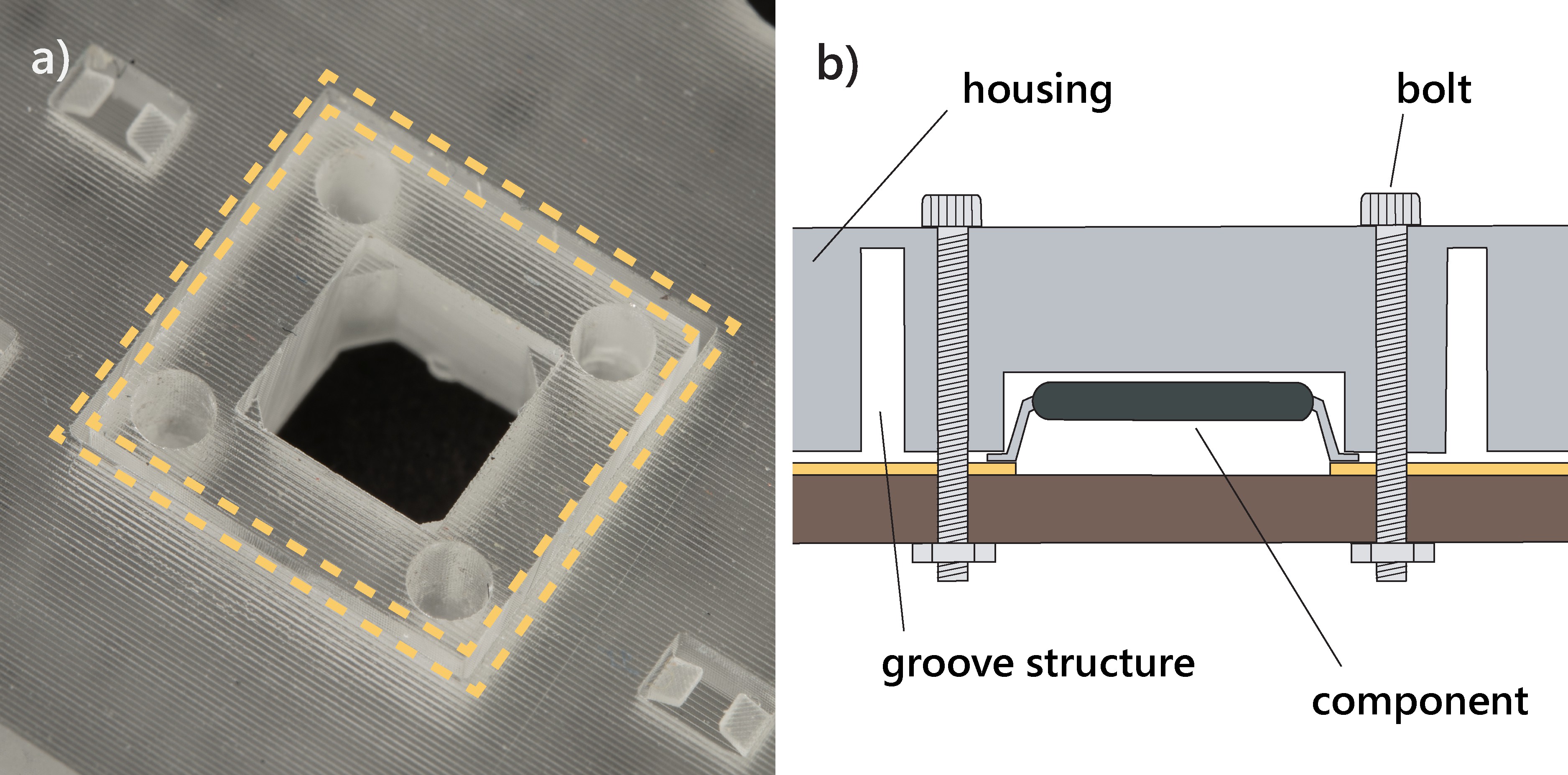}
  \caption{Groove structure: a) around the IC cavity (highlighted), b) illustration of the side sectional view of grooves surrounding a component.}
  \Description{Figure 8: This figure contains two images labeled in 'a' and 'b' showing the groove structure of the housing. Image 'a' shows the grooves location around the IC cavity. Image 'b' illustrates the side sectional view of grooves surrounding a component. }
  \label{fig:chipGroove}
\end{figure}

For both Type 1 and Type 2 cavities, we employ the bolting technique introduced in Section~\ref{bolting} to generate localized downward pressure. 
A strong force applied to the corners of the ICs by the bolts may cause local deformation in the housing structure. This could affect neighboring small two-terminal components when multiple such components are required for a PCB design. 
This issue can be alleviated by carving a groove in the housing surrounding each component that requires bolting (Figure \ref{fig:chipGroove}).
The groove's depth is set so that there is only a \SI{0.3}{\milli\meter} thick of material connecting the component housing to the main housing.
The \SI{0.3}{\milli\meter}-thick connection will deform to absorb energy and prevent housing deformation from propagating to the neighboring components.
The entire structure of the 3D housing remains as one piece during printing.

\subsubsection{Custom form factor components}

For SMD components that do not come in standard packaging, we individually design a customized cavity for each one. 
All these designs are tested for proper electrical connectivity and are stored in a component library.

\subsection{Fabrication}
All housings are 3D printed with a desktop LCD resin printer~\cite{elegoomars}. 
Since the cavity design includes miniature tabs that act as a deformable buffer (Section~\ref{0805}), the standard resin is too brittle for this purpose. 
Following the printing guidelines for flexible resin~\cite{fabhydro} and through experimentation, we empirically found that a mixture of tough UV resin~\cite{toughResin} and flexible UV resin~\cite{resionedatasheet} in a 3:2 weight ratio produces reliable printing results. This mix allows the detailed tab features to function as intended.

\subsection{Design Workflow} \label{workflow}
The SolderlessPCB housings in our work were designed using Autodesk Fusion 360, which offers integrated PCB design and 3D modeling features. 
To facilitate the design process, we developed a custom IC component library that contains the bolt hole locations and the 3D cavity designs co-existing with each IC's symbol and footprint. The library is open-sourced and can be found at: \textit{\url{https://github.com/zyyan20h/solderlessPCB}}.
Below, we briefly explain the PCB and housing design workflow using the IC component library.
\begin{figure}[h]
  
  \includegraphics[width=\columnwidth]{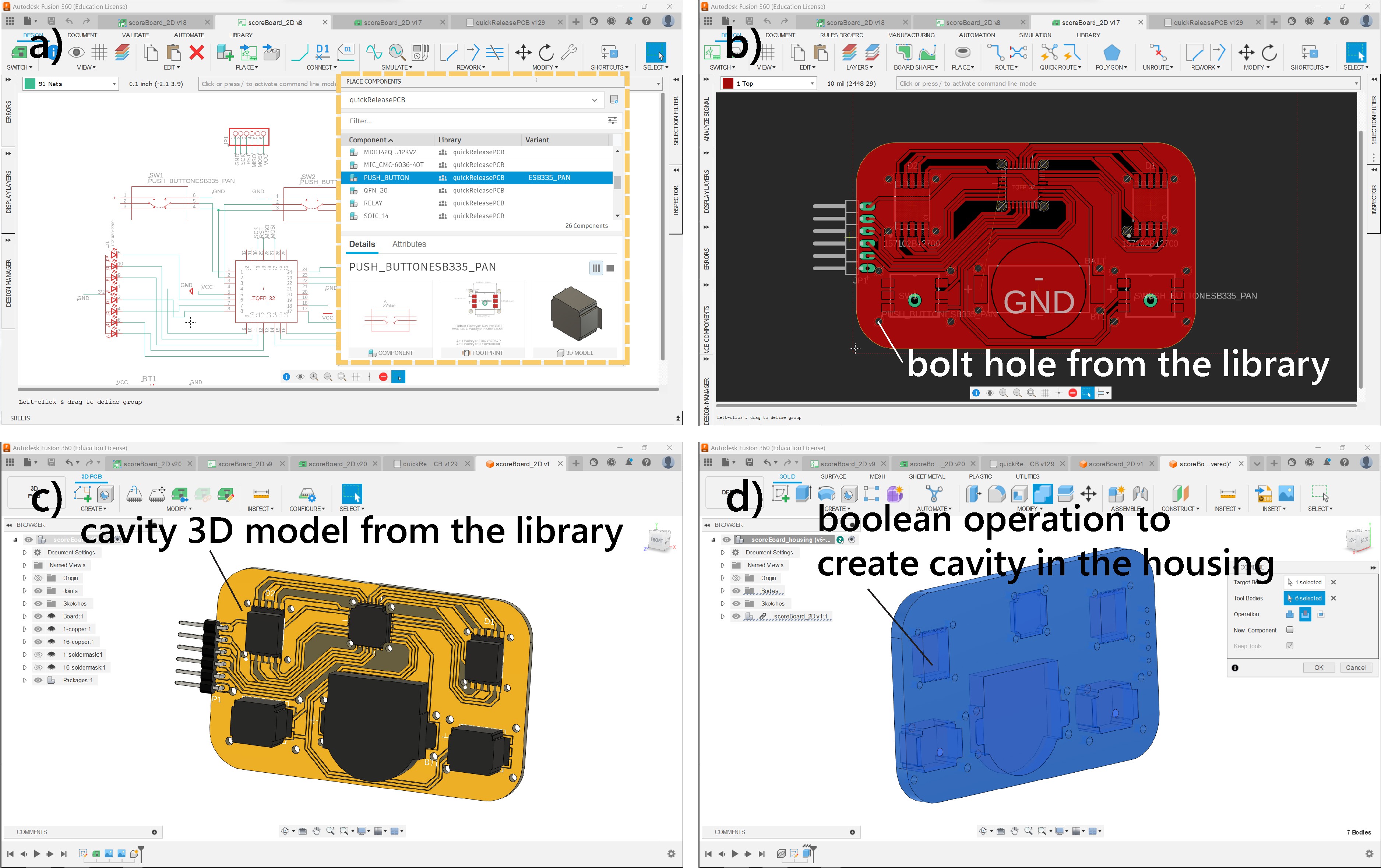}
  \caption{Design workflow: a) import component from the library and build schematic, b) route the PCB while working with bolt holes linked from the library footprint, c) push the PCB design to 3D automatically with 3D cavity models linked from the library, d) use Boolean operation to create cavities in the housing to accommodate components.}
  \Description{Figure 9: This figure contains four screenshots labeled in 'a', 'b', 'c', and 'd' showing the SolderlessPCB housings design workflow using Autodesk Fusion 360. Screenshot 'a' shows the importing component process from the library and build a schematic. Screenshot 'b' shows the PCB routing while working with bolt holes linked from the library footprint. Screenshot 'c' shows the PCB design to a 3D model including 3D cavity models linked from the library. Screenshot 'd' shows the boolean operation to create cavities in the housing to accommodate components.}
  \label{fig:workflow}
\end{figure}

As shown in Figure \ref{fig:workflow}a, the design begins with the standard PCB design pipeline, from schematic to 2D layout. 
In this stage, the user completes the circuit schematic by placing and connecting symbols of electronic components from our custom IC library onto the canvas. 
When transitioning from the schematic design to the PCB design (Figure \ref{fig:workflow}b), the bolt holes associated with each IC component are automatically rendered on the board design canvas.
The cavity models for each IC are also automatically loaded into the 3D modeling canvas, as shown in Figure \ref{fig:workflow}c. 
To create a SolderlessPCB housing, the user simply extrudes a 3D shape based on the PCB board profile to the height of the intended housing thickness. 
A Boolean operation between the extruded shape and the 3D cavity models of the ICs results in the final housing design, as shown in Figure \ref{fig:workflow}d.
Once the design is finalized, the PCB design file is sent to a CNC milling machine, and the housing design can be exported and sliced for 3D printing.

\section{Characterization and Validation}
In this section, we present a set of experiments to characterize our SolderlessPCB approach, including design characterization, electrical validation, and durability testing.

\subsection{Design Parameter Characterization}\label{design parameter chara}
\subsubsection{Bolts allocation}\label{hole distance}

\begin{figure}[h]
  
  \includegraphics[width=\columnwidth]{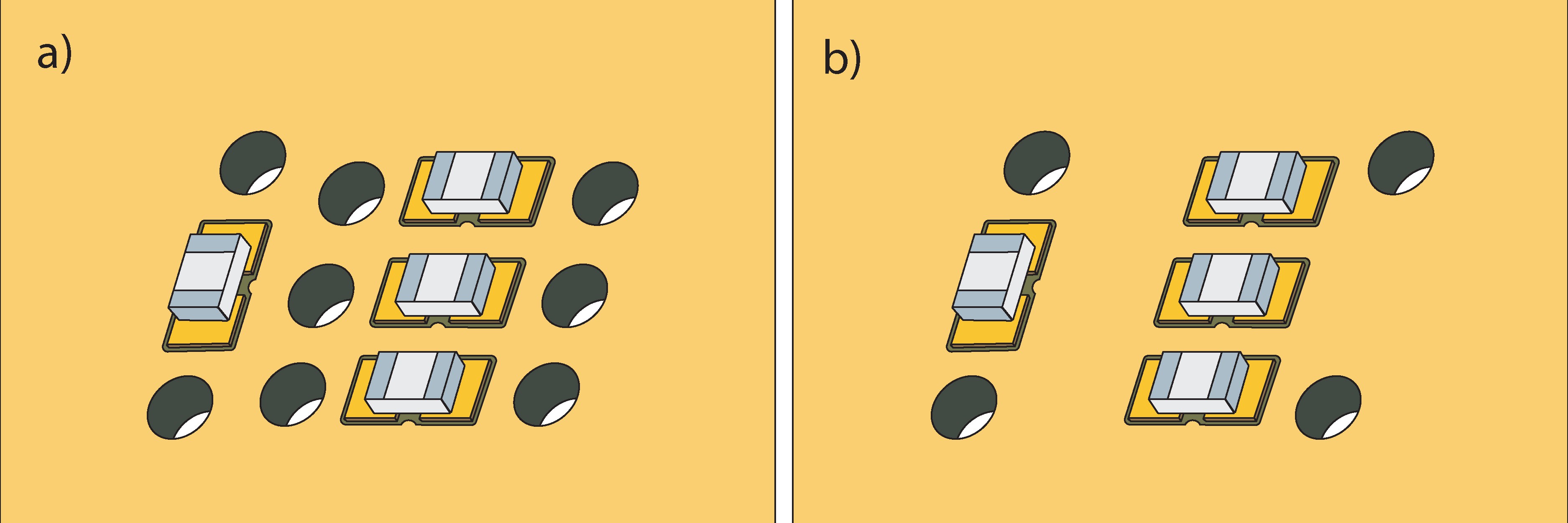}
  \caption{Bolt hole allocation for two-terminal components: a) assigning bolt holes to all closely-spaced two-terminal components may lead to an excess of bolts, b) achieving the same connection quality using only half the number of bolts.}
  \Description{Figure 10: This figure contains two illustrations labeled 'a' and 'b' showing the bolt hold allocation for two-terminal components. Illustration 'a' shows assigning bolt holes to closely-spaced two-terminal components may lead to an excess of bolts. Illustration 'b' shows achieving the same connection quality using only half the number of bolts. }
  \label{fig:lessHoles}
\end{figure}

In the IC component library, we have incorporated pre-allocated bolt holes into each IC footprint, and the same approach can also be extended to two-terminal components. However, it's a common practice in PCB design to place two-terminal components closely together, as illustrated in Figure \ref{fig:lessHoles}. In such scenarios, a set of predefined bolt holes for each two-terminal component may be unnecessary and could simply increase the manual assembly time. Instead, it is possible for a group of closely placed two-terminal components to share bolts (Figure \ref{fig:lessHoles}b). Here, we report on an experiment that determines the maximum allowable distance between neighboring bolts for a specific thickness of 3D-printed housing. The results of this experiment can serve as guidelines for the placement of bolts in two-terminal components that do not have pre-allocated holes.

\begin{figure}[h]
  
  \includegraphics[width=\columnwidth]{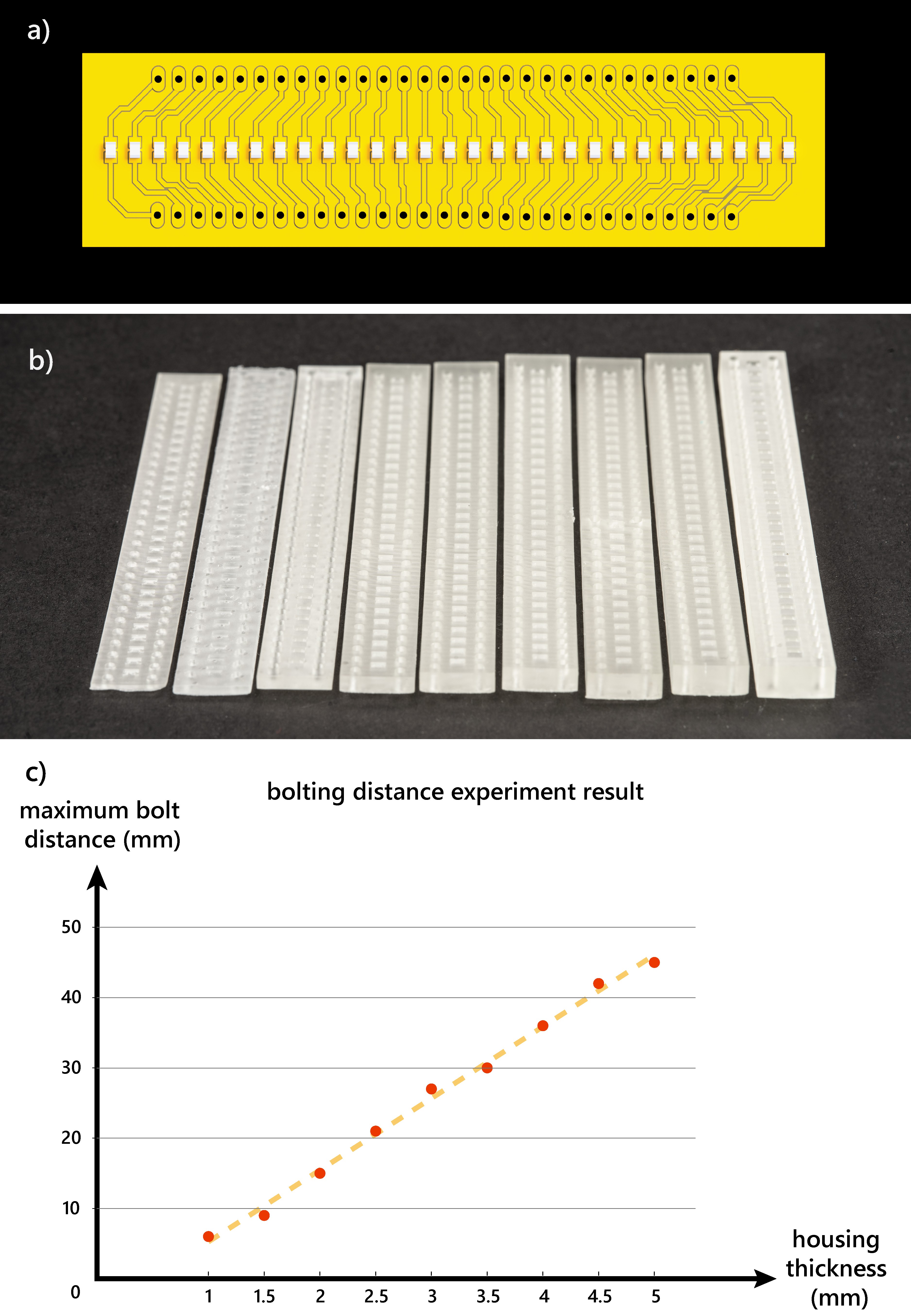}
  \caption{Bolting distance experiment: a) rendering of the PCB for the experiment; b) housings in different thicknesses; c) graph showing the maximum bolting distance versus housing thickness.}
  \Description{Figure 11: This figure contains three images labeled in 'a', 'b', and 'c' showing the bolting distance experiment and results. Image 'a' illustrates the experimental PCB design. Image 'b' shows the photo of sample housings in different thicknesses. image 'c' contains a graph showing the maximum bolting distance versus housing thickness. The maximum distance between bolt holes monotonically increases with the thickness of the housing.}
  \label{fig:distanceExp}
\end{figure}

As shown in Figure~\ref{fig:distanceExp}a, we prepared a \SI{23}{\milli\meter} by \SI{91}{\milli\meter} PCB with a pair of bolt holes every \SI{3}{\milli\meter} along the longitudinal axis. Between each pair of bolt holes, there is space for a 0805 two-terminal component. In total, the PCB can host 29 0805 SMD resistors.
We also prepared nine 3D printed housings the size of the PCB, with the thickness ranging from \SI{1}{\milli\meter} to \SI{5}{\milli\meter}, at intervals of \SI{0.5}{\milli\meter}. 

In the experiment, we first placed all 29 0805 0-ohm resistors into one of the 3D printed housings. We then bolted the housing to the PCB, installing one pair of bolts at the very left end of the PCB, and another pair of bolts starting from the very right end and moving towards the first pair. While moving this second pair of bolts, we measured the conductivity of all resistors located between them and the left pair, until all resistors in between showed continuous connection. The distance between the two pairs is the maximum distance between bolts that can ensure reliable conductivity for the given thickness of the housing. 

We repeated the same experiment for all nine housings, with each housing tested three times. As shown in the graph (Figure~\ref{fig:distanceExp}c), the maximum distance between bolt holes monotonically increases with the thickness of the housing. This experiment enables us to determine the maximum distance between bolts for each housing thickness. For example, for a 3D printed housing with a thickness of 3 mm, as long as the distance between bolts is less than 27 mm, we can ensure that the electronic components in between have a reliable connection to the base PCB. All examples presented in this paper adhere to this design rule.

\subsubsection{Tab dimensions} \label{tabDim}

For two-terminal components, the flexible tab structure must be sufficiently pliable to absorb energy, thereby preventing the housing from bending. 
At the same time, it must exert enough pressure on the components to ensure good conductivity without causing breakage.
We experimented with different geometric parameters and converged on the following empirical formula:

$$T = t;
\alpha = \ang{30};
W = 0.6 \times w;
L = 0.45 \times l;
H = t + \SI{0.1}{\milli\meter};
$$

Where \textit{$T$} represents tab thickness, \textit{$\alpha$} is the tab's slope angle, \textit{$W$} and \textit{$L$} denotes width and length, and \textit{$H$} indicates the height with respect to the housing surface. These parameters are determined by the specifications of a two-terminal component, where \textit{t, w, l} represent the rated component thickness, width, and length, respectively.

\begin{figure}[h]
  
  \includegraphics[width=\columnwidth]{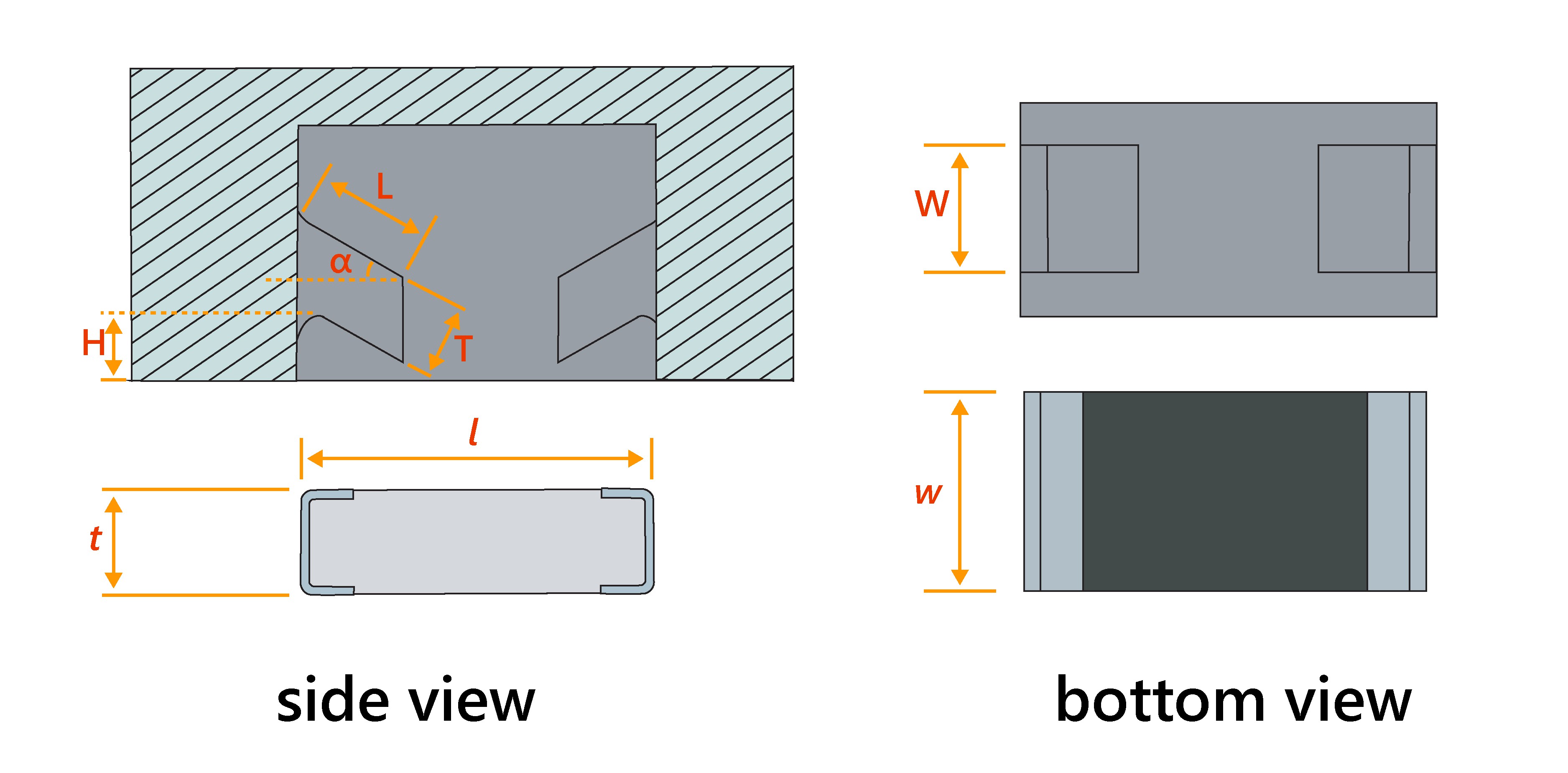}
  \caption{Tab dimensions with respect to component.}
  \Description{Figure 12: This figure shows the dimensions with respect to component. }
  \label{fig:tabDim}
\end{figure}

We report that, using the aforementioned parameters and setup, we can reliably print housings for two-terminal components the size of or larger than the 0603 series. For the 0603 package, the tab has a length of \SI{0.69}{\milli\meter}, a width of \SI{0.48}{\milli\meter}, and a thickness of \SI{0.45}{\milli\meter}.

\subsubsection{IC cavity validation} \label{validation}
To validate our housing cavity design for ICs, we developed a simple testing circuit incorporating ATtiny microcontrollers. We specifically selected three types of ICs: ATtiny85, ATtiny84A, and ATtiny828, chosen for their identical architecture, which enables them to run the same testing code. This selection enables us to test five different yet commonly used SMD packages: the ATtiny85 is available in BGA, SOIC, and TSSOP; the ATtiny84A in a QFN package, and the ATtiny828 in a TQFP package. Among the five different packages, the TSSOP has the minimal pitch between pins at \SI{0.6}{\milli\meter}; the BGA has the smallest contacting pad area of \SI{0.0625}{\square\milli\meter}. 

\begin{figure}[h]
  
  \includegraphics[width=\columnwidth]{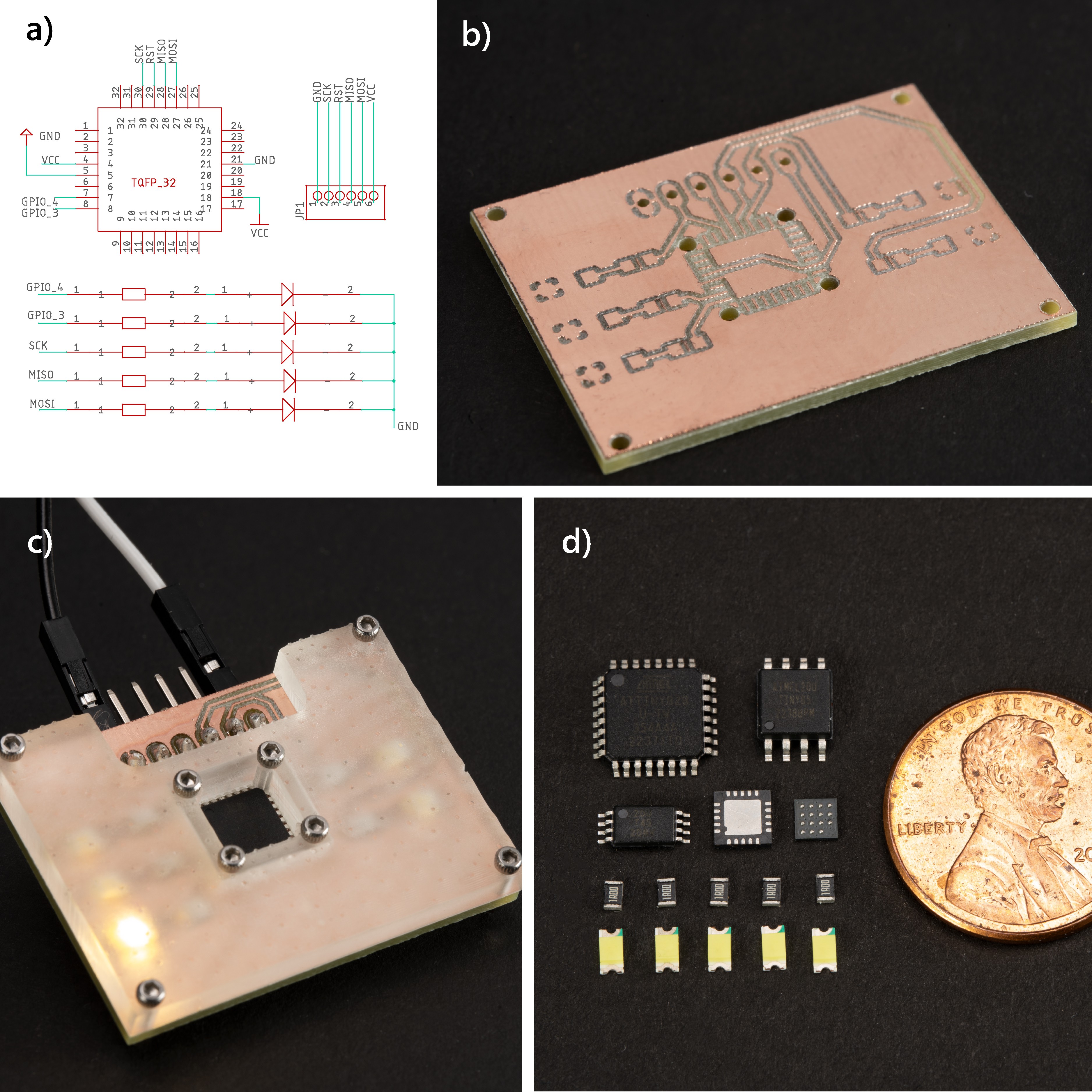}
  \caption{IC validation: a) schematic used for IC package validation, b) example circuit board used for validation (TQFP-32), c) example PCB assembly (TQFP-32), d) all packages and components used during validation.}
  \Description{Figure 13: This figure contains four images labeled in 'a', 'b', 'c', and 'd' showing the IC validations. Image 'a' shows a schematic used for IC package validation. Image 'b' shows an example circuit board used for validation (TQFP-32). Image 'c' shows an example PCB assembly (TQFP-32). Image 'd' shows all packages and components used during validation.}
  \label{fig:chipPackages}
\end{figure}

Our testing circuit comprises five LEDs, each connected to one of the five GPIO pins of the ATtiny ICs. Every LED features a 1206 footprint and is serially connected to a \SI{100}{\ohm} resistor in a 0805 package. Power is supplied to the circuit through two soldered through-hole pin headers. To test the housing cavity design, we uploaded code that controls the sequential flashing of each LED, as shown in Figure \ref{fig:chipPackages} and the accompanying video. We confirmed that all five packages establish reliable mechanical contact with the PCB baseboards. As a result, they effectively control the LED blinking, free from any glitches or ghost connections.

\subsection{Electrical Characterization}\label{electric characterization}
As the SolderlessPCB approach alters the method of bonding electronic components to the PCB board, we conducted three sets of experiments to understand how it might impact specific circuit electric characteristics. Specifically, we focused on examining three aspects of a circuit: resistance, impedance, and high-frequency signal energy loss, as these are critical for circuits with AC and DC functionalities.

\subsubsection{Resistance}\label{resistance}
\begin{figure}[h]
  
  \includegraphics[width=\columnwidth]{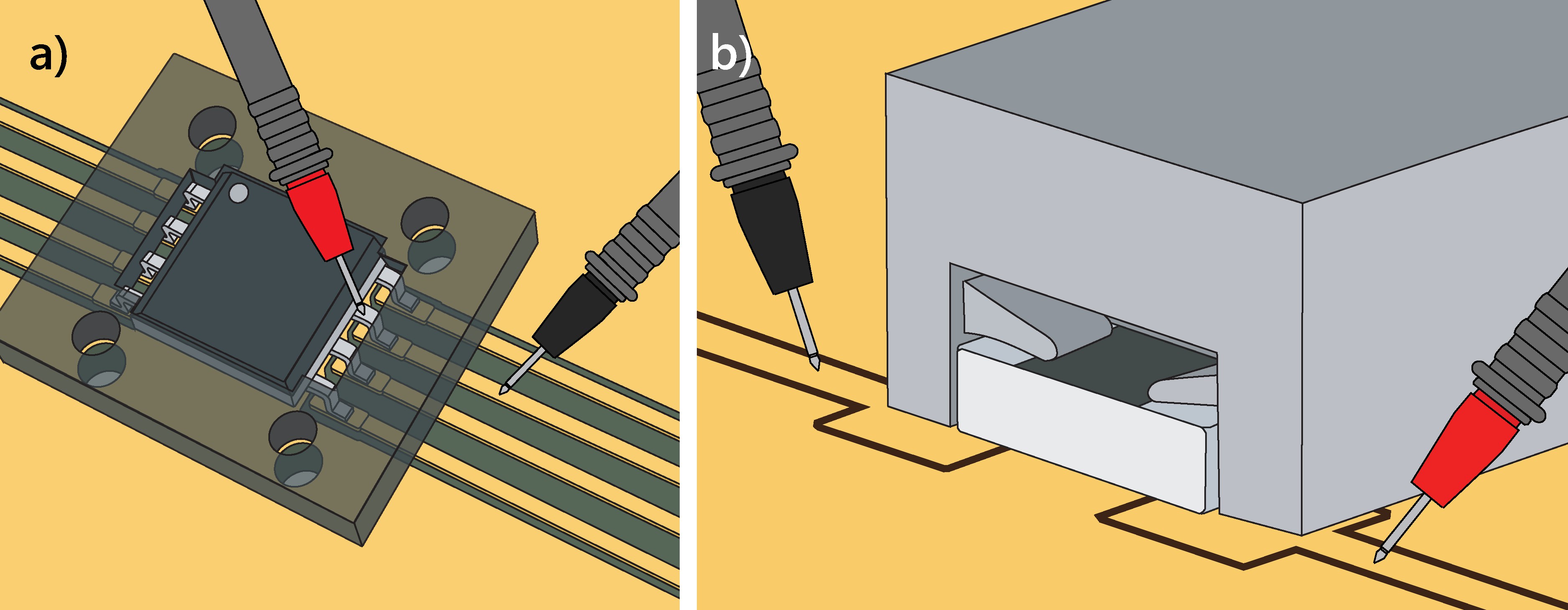}
  \caption{Measurement procedure: a) the resistance measured at IC pins, b) the resistance measured for two-terminal components.}
  \Description{Figure 14: This figure contains two illustrations, labeled in 'a' and 'b', showing the measurement procedure. Illustration 'a' shows the resistance measured at ICs with two probes. One probe lead touches the IC's lead, the other probe lead touches contacting trace. Illustration 'b' shows the resistance measured for two-terminal components. }
  \label{fig:probing}
\end{figure}

We measured 177 connection points created by SolderlessPCB across various packages of SMD components. 
For each measurement, we produced custom PCB boards and housings, enabling us to place two measurement probe leads: one at the IC's metal lead, and the other at the contacting trace for ICs, or alternatively, at each side of the contacting traces for two-terminal components. 
Figure \ref{fig:probing} illustrates our measuring setup. 
We measured the resistance using a Keysight 3446SECU digital multimeter and reported an average resistance of \SI{0.46}{\ohm} with a standard deviation of \SI{0.139}{\ohm}. The extra resistance should not affect the performance of common DC circuits when the electrical load is significantly larger than the connection point resistance.
Note that the resistance data does not include connection points for any BGA package due to the lack of exposed conductors available for direct measurement, however, all prototypes made with UFBGA-15 package fulfill their functionalities.

\subsubsection{High-frequency signal energy loss}

Besides resistance, the quality of the electrical connection also affects signal transmission power, especially at higher frequencies ~\cite{chen2013signal,sharawi2004practical}.

\begin{figure}[h]
  
  \includegraphics[width=\columnwidth]{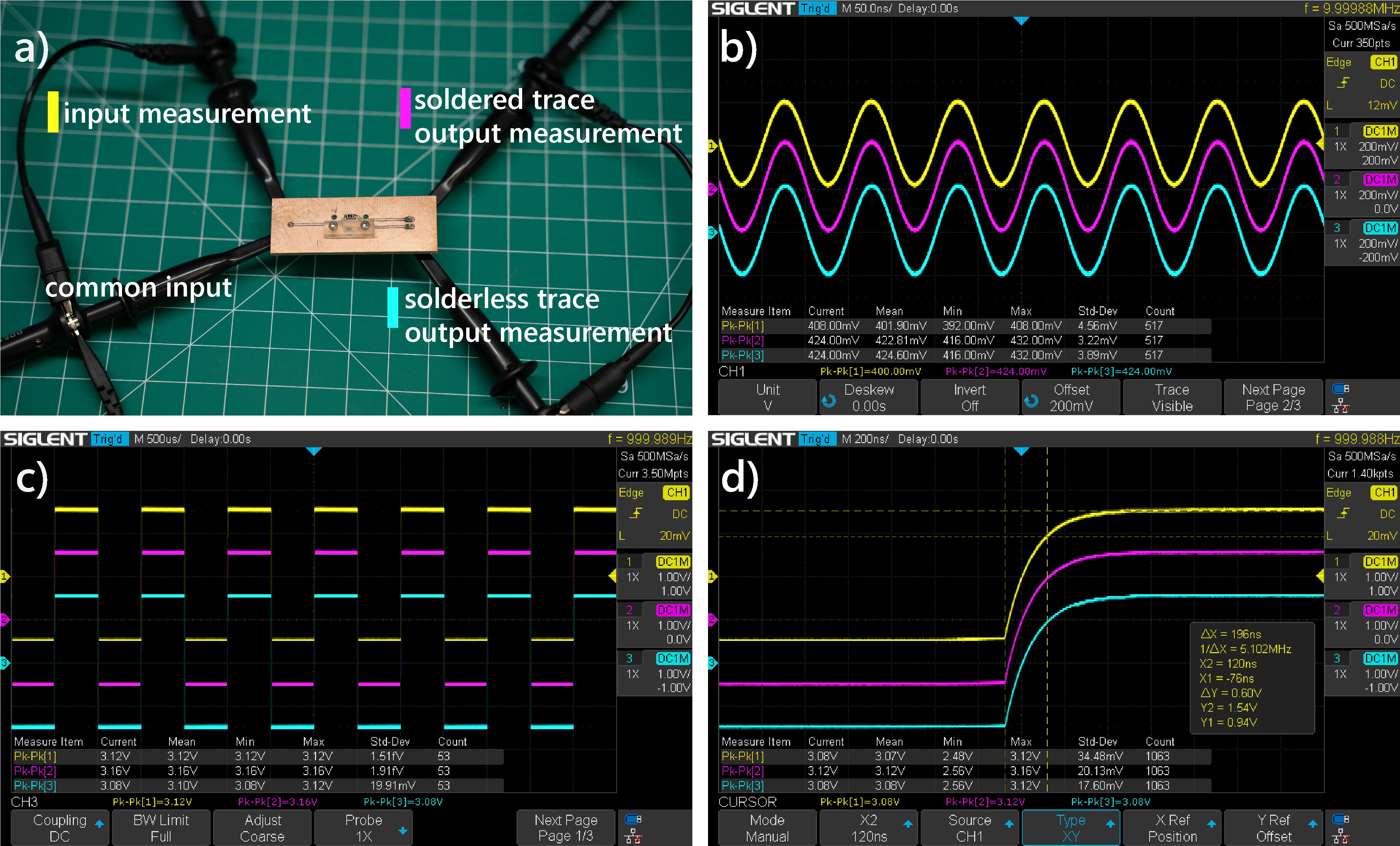}
  \caption{Waveform results: a) the PCB used for the waveform experiment, b) the sinusoidal wave measured from both input and output terminals, c) the quasi-square wave measured from both input and output terminals, d) zoomed-in view in the time domain of the quasi-square wave measured from all terminals.}
  \Description{Figure 15: This figure contains four images labeled in 'a', 'b', 'c', and 'd' showing the waveform experiment setup and results. Image 'a' shows the PCB used for the waveform experiment. Image 'b' shows the results of the sinusoidal wave measured from both input and output terminals. Image 'c' shows the quasi-square wave measured from both input and output terminals. Image 'd' shows the zoomed-in view in the time domain of the quasi-square wave measured from all terminals. The results of the waveform show there is no waveform attenuation through either the soldered or solderless circuit.  }
  \label{fig:wave}
\end{figure}

To understand how SolderlessPCB preserves waveforms across different frequency ranges, we designed a test PCB equipped with two identical traces sharing the same input pin header. One trace includes a soldered \SI{0}{\ohm} resistor, while the other features a \SI{0}{\ohm} resistor assembled using the SolderlessPCB approach (Figure \ref{fig:wave}a). We supplied both a sinusoidal wave and a quasi-square wave to the common pin header using a Function Generator BK PRECISION 4053, with frequencies ranging from \SI{100}{\hertz} to \SI{10}{\mega\hertz} (the equipment's maximum frequency). We then output the waveform from both output terminals and the input terminal as a benchmark, using an Oscilloscope SIGLENT SDS 1104X-E (Figure \ref{fig:wave}b, c, and d). The results indicate no waveform attenuation through either the soldered or solderless circuit. For example, in the output graphs shown in Figure \ref{fig:wave}, both signals–the soldered trace (pink curve) and the solderless trace (blue curve)–exhibit the same amplitude level as the input signal (yellow curve), for both the sine wave and the zoomed-in view of the square wave.

\begin{figure}[h]
  
  \includegraphics[width=\columnwidth]{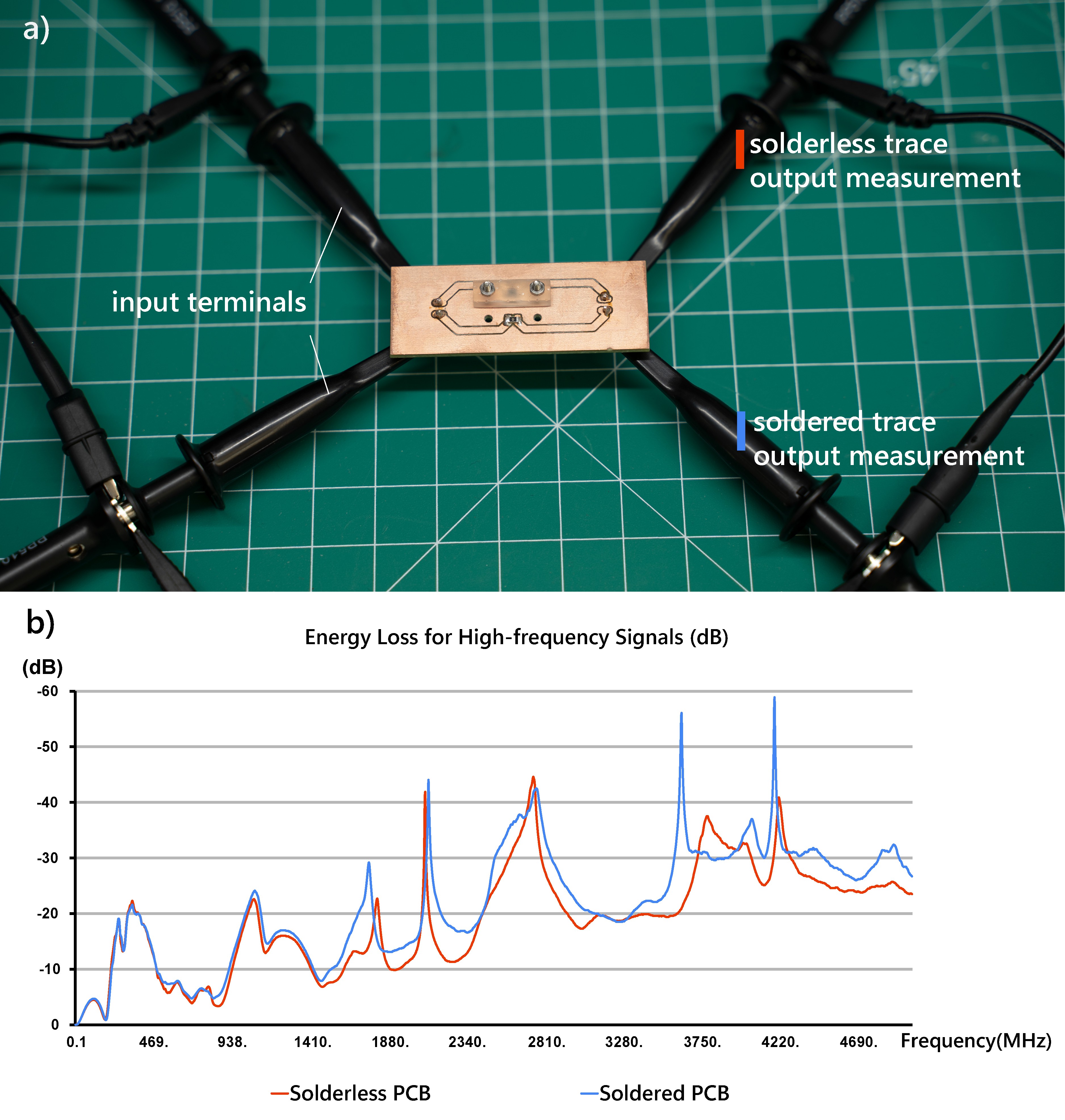}
  \caption{Signal energy loss for solderless and soldered PCB. a) the PCB used for the waveform experiment, b) the energy loss curve (dB) in the frequency domain.}
  \Description{Figure 16: This figure contains two images labeled in 'a' and 'b' showing the experiment setup for the signal energy loss and results. Image 'a' shows the experiment setup for the signal energy loss. Image 'b' shows the graph of the testing result.}
  \label{fig:signal_loss}
\end{figure}

To further assess the power loss of the high-frequency signals through mechanical connection points compared to soldered connections, we measured the energy loss between input and output signals using two \SI{0}{\ohm} resistors. Each resistor was assembled to close the circuit using either a soldered connection or the SolderlessPCB approach. To minimize energy loss on the trace itself, we redesigned the PCB with 65 mil traces (Figure~\ref{fig:signal_loss}a). We also used individual input pins for each trace to prevent uneven splitting of signal energy at a shared input pin. For accurate power loss data, the test PCB was connected to a Vector Network Analyzer Keysight E5071C and fed with a pure sinusoidal wave signal within a range of \SI{100}{\kilo\hertz} to \SI{5}{\giga\hertz}. The average signal energy loss across the full testing frequency was \SI{-18.96}{\decibel} for SolderlessPCB and \SI{-21.61}{\decibel} for the soldered PCB, with higher frequencies exhibiting greater energy loss in both cases (Figure~\ref{fig:signal_loss}b).

The results of our experiment show that the SolderlessPCB is comparable to the soldered PCB even for high-frequency signal transmission.

\subsubsection{Impedance circuit reproducibility}
Impedance is crucial in prototyping applications like radio, Bluetooth, and WiFi antennas, where impedance matching circuits are often developed through an iterative process \cite{9646246}. In such scenarios, minimizing impedance variance at the connection interfaces between components and the PCB during assembly and disassembly is preferable. To assess impedance variance through multiple (dis)assembly iterations with SolderlessPCB, we designed and fabricated two identical PCBs carrying a pi-network circuit, commonly used for impedance matching \cite{6675068}. We assembled the same set of components, both soldered and solderless, on these boards. Using a Vector Network Analyzer Keysight E5071C, we measured the impedance values for both PCBs across five (dis)assembly iterations. A pure sinusoidal wave signal ranging from \SI{100}{\kilo\hertz} to \SI{5}{\giga\hertz} was supplied to the input end. It should be noted that frequencies below \SI{100}{\kilo\hertz} were not tested, as they are not relevant for impedance circuit applications \cite{chen2013signal, sharawi2004practical}.

\begin{figure}[h]
  
  \includegraphics[width=\columnwidth]{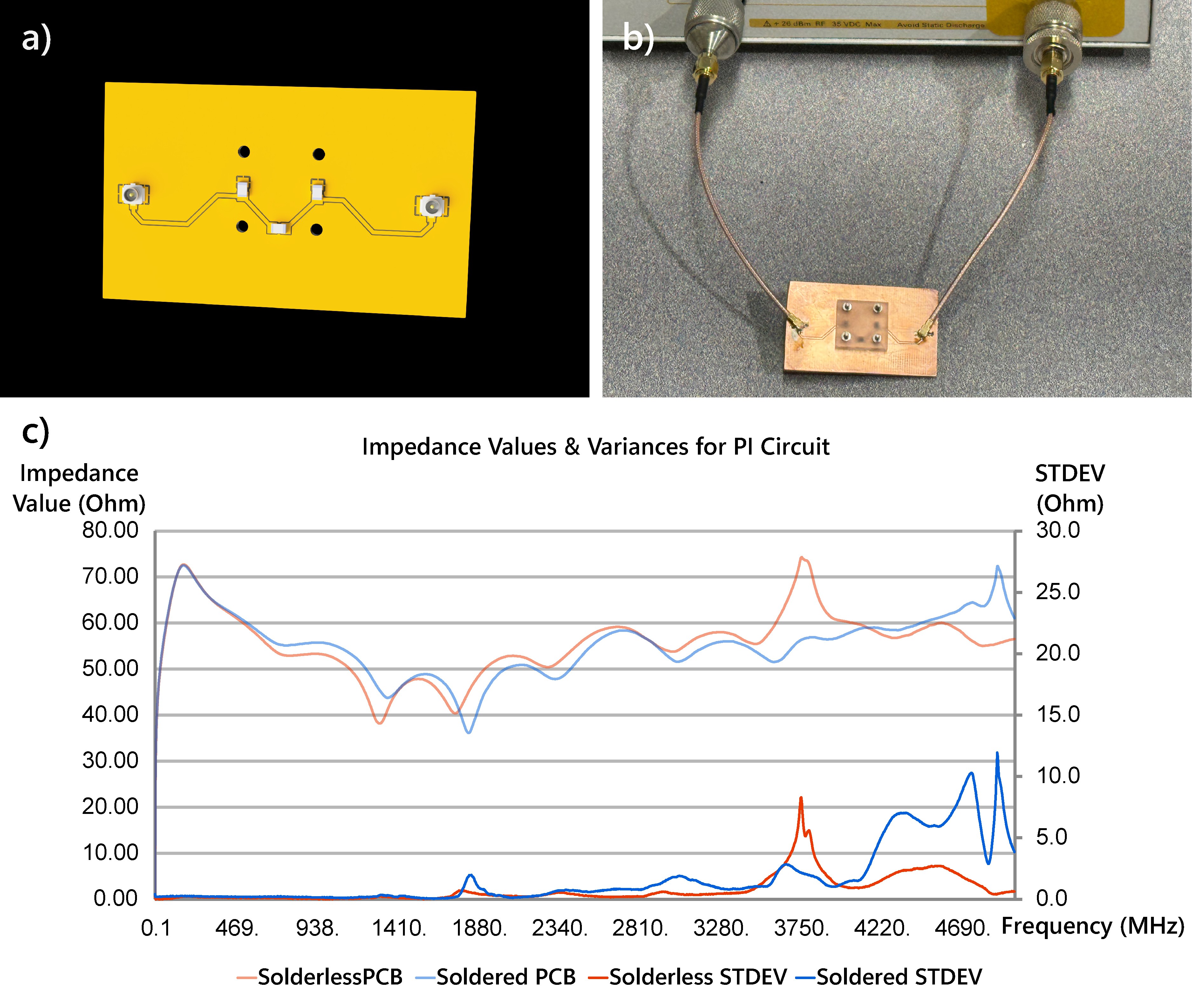}
  \caption{Impedance values between the solderless and soldered PI PCB: a) rendering of the PCB used in the impedance experiment, b) the PCB connection and experiment setup, c) the impedance values and variances (Ohm) in the frequency domain. }
  \Description{Figure 17: This figure contains three images labeled in 'a', 'b', and 'c' showing the experiment setup for the impedance experiment and results. Image 'a' shows the PCB used in the experiment. Image 'b' shows the experiment setup. Image 'c' shows the graph of the testing result.}
  \label{fig:imp_pi}
\end{figure}

We report the results for both the soldered and solderless PCBs, along with their variances, in Figure~\ref{fig:imp_pi}. As shown in the figure, the impedance variance for solderless PCBs across different assembly iterations remains on par with that of the soldered ones. The results indicate that SolderlessPCB can be used for prototyping impedance matching circuits and other similar applications.

\subsection{Reliability Test} \label{gforce}
We performed additional tests to understand the reliability of SolderlessPCB.

\subsubsection{Housing reliability}
All prototypes described in Section~\ref{design parameter chara} and Section~\ref{walkthroughs}, including those with both IC cavity and tab housings, remained fully functional three months post-assembly. This demonstrates that our SolderlessPCB methods can be reliably used for prototyping purposes.

\begin{figure}[h]
  
  \includegraphics[width=\columnwidth]{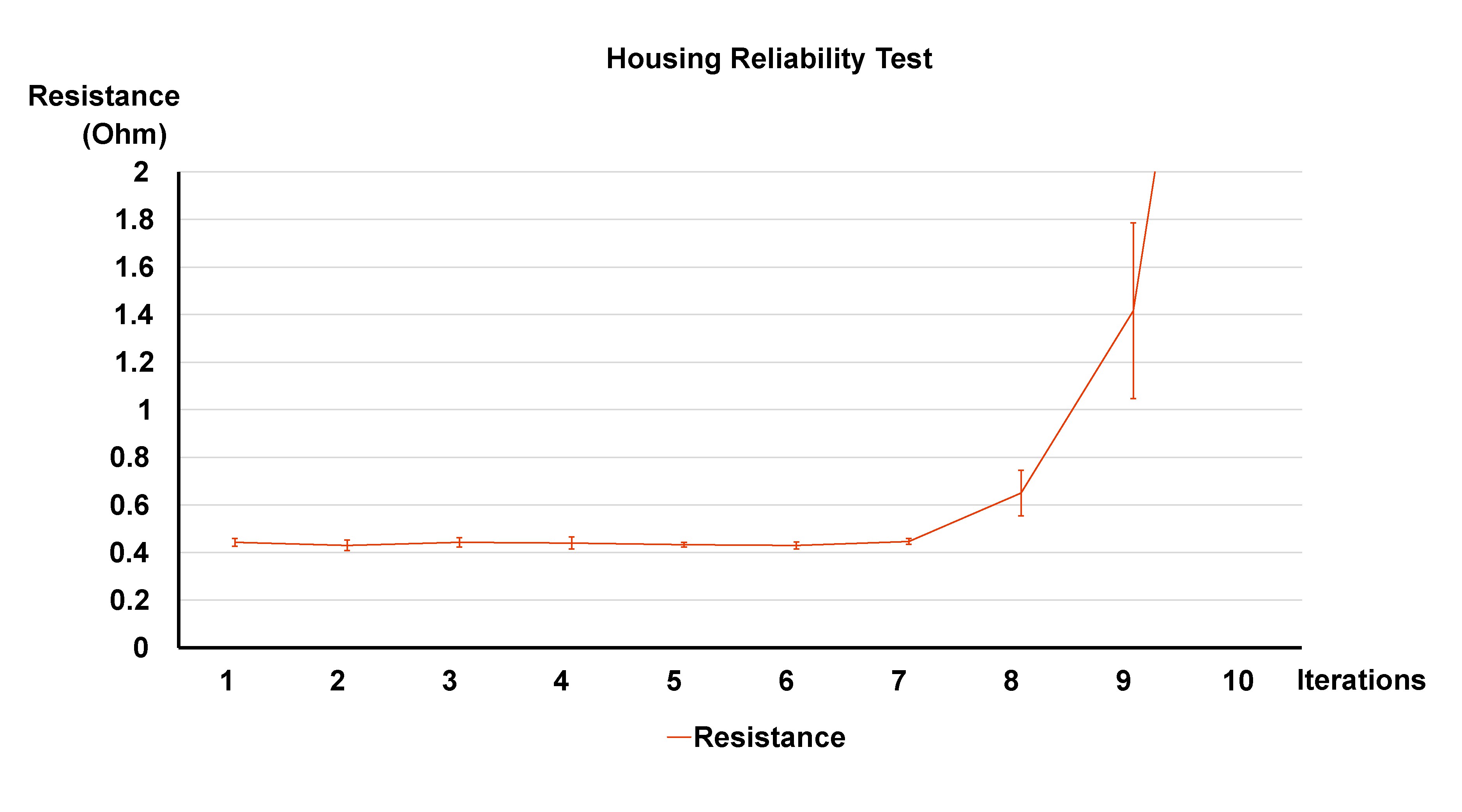}
  \caption{Resistance at contact pads after each disassembly-assembly iteration.}
  \Description{Figure 18: This figure shows the experiment results of resistance at contact pads after each disassembly-assembly iteration.}
  \label{fig:dura}
\end{figure}

As one potential application of SolderlessPCB is to facilitate the replacement of defective SMD components with new ones, we further investigated the housing reliability for component switching. Our focus was specifically on the tab structures of the two-terminal component housing, which are printed on a sub-millimeter scale and can be vulnerable to fatigue during the re-installation of a housing.
To test the reliability of the printed tab structures, we repeatedly assembled and disassembled housings of three 0805 \SI{0}{\ohm} resistors.
For each iteration, we unbolted the housing and took out the resistors, and reassembled them from scratch.
Using the same method as in Section ~\ref{resistance}, we measured the resistance values at the connection points after each assembly. 
We report the average resistance value across ten disassembly-assembly iterations in Figure \ref{fig:dura}.
The resistance value remains consistent for the initial assembly cycles but begins to increase after the seventh iteration, indicating a failure of the tabs to exert adequate pressure beyond this point. We conclude that SolderlessPCB can be reliably used for component switching up to seven times before a new housing is required. 

\subsubsection{Drop test}
We conducted a drop test to investigate the robustness of SolderlessPCB. Specifically, we dropped the PCB assembly described in Section \ref{validation} from heights ranging from \SI{0.3}{\meter} to \SI{2}{\meter}, at \SI{0.3}{\meter} intervals. We conducted three drops from each height; the PCB assembly remained fully functional after the entire series of drop tests. For an additional stress test, we dropped the same PCB assembly from a height of \SI{6}{\meter} three more times. The PCB assembly continued to function properly. Our accompanying video documented one of these tests.

\section{Scenarios and Walkthroughs}\label{walkthroughs}
In this section, we present two walkthrough examples to demonstrate: 1) how SolderlessPCB enables designers to create prototypes without soldering; 2) how it allows for the reuse of old electronic components in new projects; and 3) how SolderlessPCB aids in iterative prototyping practices by requiring fewer new electronic components.

\subsection{From Kitchen Timer to Foosball Scoreboard: Reusing SMD Components for a New Project}
\subsubsection{Building a timer with the SolderlessPCB approach.} \label{timer}
Alex is a maker and a designer. 
She recently moved to a new apartment and wanted to take up cooking as a new hobby. 
While the kitchen was well-equipped, it lacked one essential tool for Alex: a kitchen timer. Rather than using a smartphone, which could get greasy during cooking, or purchasing a brand-new kitchen timer, which might take time to be delivered, Alex decided to build her own timer using electronic parts she had saved previously.

\begin{figure}[h]
  
  \includegraphics[width=\columnwidth]{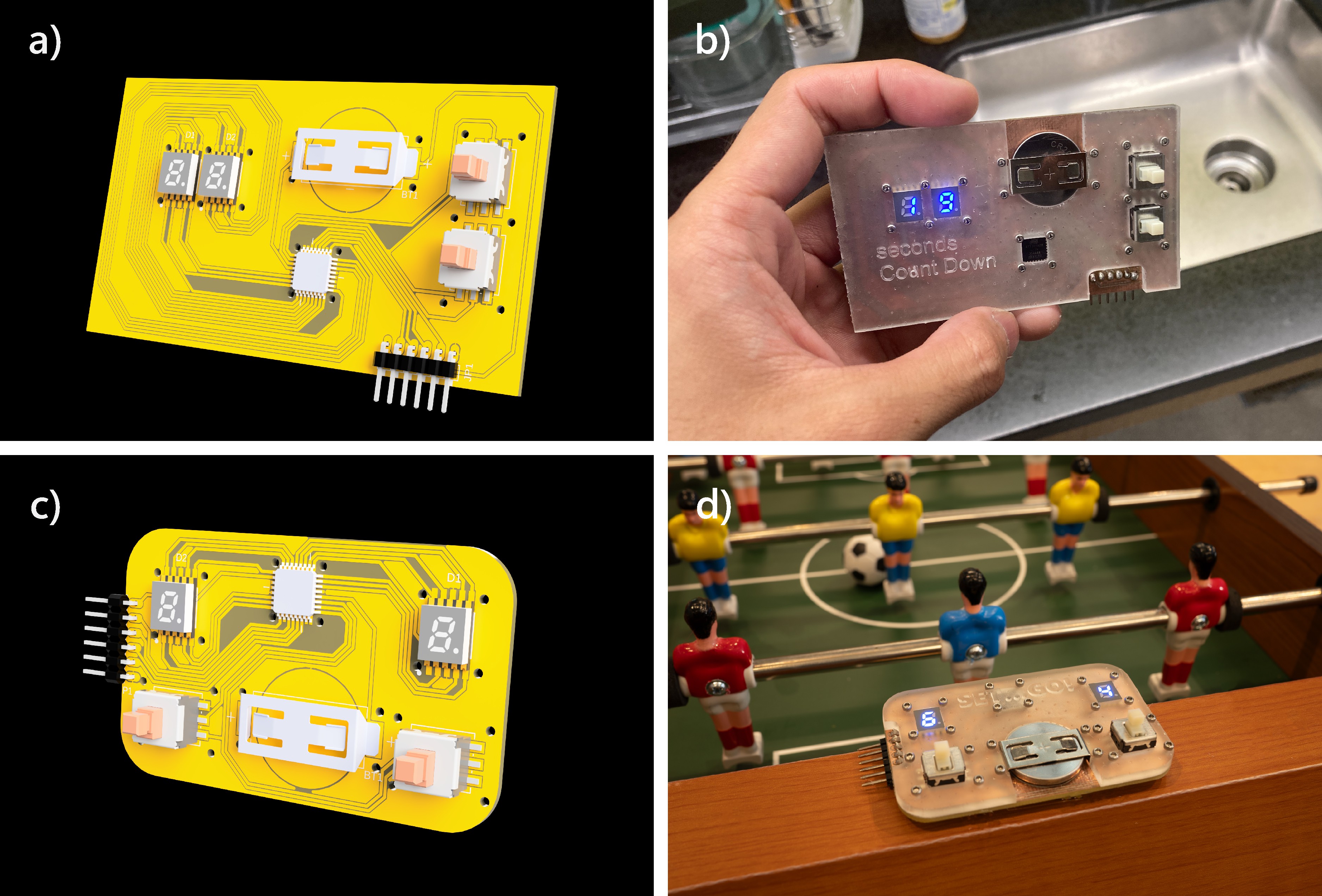}
  \caption{Kitchen timer example: a) and b) the kitchen timer was fully assembled without soldering. c) and d) when Alex decided to start a new project, she was able to fully salvage the SMD components from the old timer and repurposed them to create the scoreboard.}
  \Description{Figure 19: This figure contains four images labeled in 'a', 'b', 'c', and 'd' showing the components reuse example. Image 'a' and 'b' show the kitchen timer was fully assembled without soldering. Image 'c' and 'd' show a new project with the SMD components from the old timer and repurposed them to create the new scoreboard. }
  \label{fig:wt1}
\end{figure}

Alex's design was a PCB measuring \SI{90}{\milli\meter} by \SI{50}{\milli\meter}, featuring the following components: one ATtiny828 IC in TQFP-32 format, two 8-digit displays, two push buttons, and a battery holder (Figure \ref{fig:wt1}a). 
Adhering to the design workflow of SolderlessPCB, Alex designed the circuit and generated the corresponding 3D housing.
She fabricated the circuit using a desktop CNC machine and printed the housing with an off-the-shelf resin printer. 
Figure~\ref{fig:wt1}b showcases the completed assembly, which functions as a 60-minute timer.

\subsubsection{Reuse electronic components for a foosball scoreboard.}
Alex recently received two birthday gifts: a professional kitchen timer and a mini foosball table—the DIY timer she made earlier was no longer needed. 
However, noting all the electronic components in the DIY timer are perfectly functioning, she decided to reuse and repurpose them to create a digital scoreboard for the new foosball table.

Thanks to the SolderlessPCB approach, the kitchen timer could be easily disassembled using just a screwdriver. 
Alex then designed a new PCB for a scoreboard, utilizing all the components from the previous project. 
In the new design, the two 8-digit displays were positioned on either side, now serving to display the game scores. 

Figure~\ref{fig:wt1}d shows that the scoreboard installed on the foosball table. 
When a player scores a goal, pressing a button will update the display accordingly.

\subsection{Design Iteration on a Bristlebot}
\subsubsection{Iteration one: malfunctioning IC}
A bristlebot is a tiny robot built from a toothbrush head. 
With bristles on its bottom, it can move in arbitrary patterns or follow a predefined trajectory when vibrating. 
Alex saw it online and would like to build one as a toy for her newborn nephew.

Alex tested the circuit with a breadboard first. 
However, as a breadboard was too big for a toothbrush head, the finished toy had to have a compact design using a custom PCB. 

\begin{figure}[h]
  
  \includegraphics[width=\columnwidth]{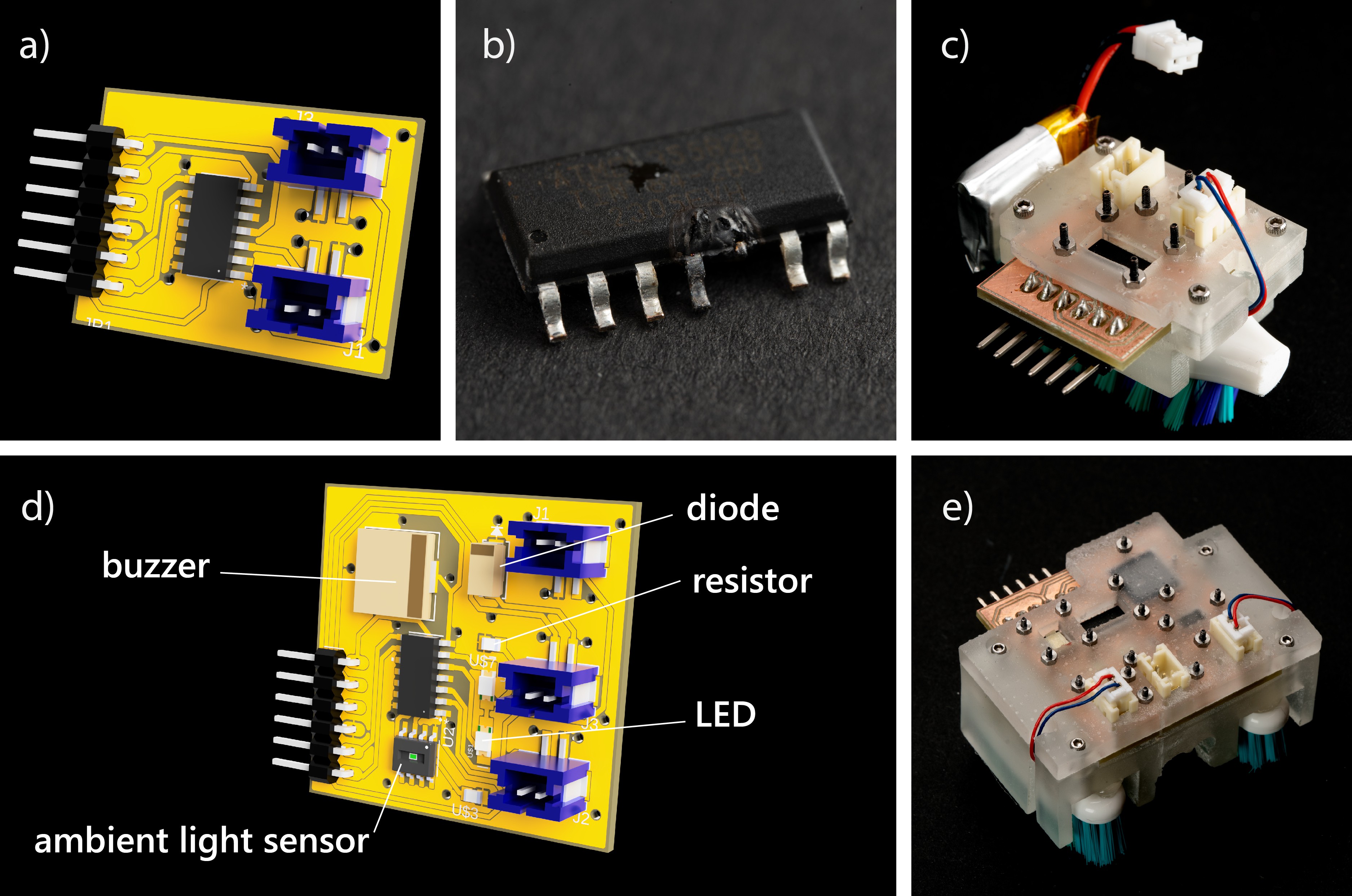}
  \caption{Bristlebot iterations: a) the first iteration PCB design, b) fried microcontroller, c) assembly of the second iteration, involving a simple replacement of the fried IC with a new one, d) the PCB design for the third iteration incorporating new sensors and actuators, e) assembly of the third iteration.}
  \Description{Figure 20: This figure contains five images labeled in 'a', 'b', 'c', 'd', and 'e' showing the Bristlebot iterations. Image 'a' shows the first iteration PCB design. Image 'b' shows a photo of a fried microcontroller. Image 'c' shows the assembly of the second iteration, involving the simple replacement of the fried IC with a new one. Image 'd' shows the PCB design for the third iteration incorporating new sensors and actuators. Image 'e' shows the assembly of the third iteration.}
  \label{fig:wt2}
\end{figure}

Like before, Alex employed the SolderlessPCB approach and manufactured her own PCB baseboard and 3D-printed housing. Figure~\ref{fig:wt2}a shows the first iteration. It featured an ATtiny84 as the microcontroller and utilized two top-open JST connectors, one reserved for a vibration motor and the other for a Li-Po battery.

Alex used a bench power supply to power the circuit for testing. 
Unfortunately, the two terminals of the supply were incorrectly connected to the board, resulting in the small IC on the circuit getting burned. Figure~\ref{fig:wt2}b shows the fried IC.

\subsubsection{Iteration two: IC replacement}
In most cases, a burned multi-pin IC can pose a problem since it is challenging to desolder from a PCB and to clean all the pins properly. 
As a result, the fried PCB and all its onboard electronic components are often disposed of directly.

In Alex's case, however, replacing the IC was much simpler. After disassembling the housing from the PCB baseboard, Alex popped out the broken IC. Noticing that the circuit traces were intact, she directly inserted a new IC into the design. This time, the circuit worked properly, and Alex had a functioning bristlebot (Figure \ref{fig:wt2}c).

\subsubsection{Iteration three: upgrading the design with new features}
Since the bristlebot lacked the capability to respond to its environment and was not controllable, Alex embarked on a third design iteration.
In this iteration, she decided to include an ambient light sensor to sense environmental light as a means to control the bot's motion. 
Additional features like an LED to indicate the bot's status and a buzzer to make sounds were also added (Figure \ref{fig:wt2}d). 
Because these extra components could add weight, Alex also upgraded the bot from one brush head to two, and with two vibration motors. 

Alex still needed to prepare a new PCB baseboard, but because she used SolderlessPCB approach, she was able to recycle all the SMD components from the second iteration. 
Alex also modified the PCB housing to include a few additional features, allowing it to clamp directly onto the two brush heads. The housing now serves two functions: it secures the SMD components to the PCB, and it also forms part of the bot's structure. Figure~\ref{fig:wt2}e shows the final design iteration.

\subsection{Additional Examples}

In this section, we showcase three additional example circuits created using SolderlessPCB, each illustrating different electronic functionalities.

\begin{figure}[h]
  
  \includegraphics[width=\columnwidth]{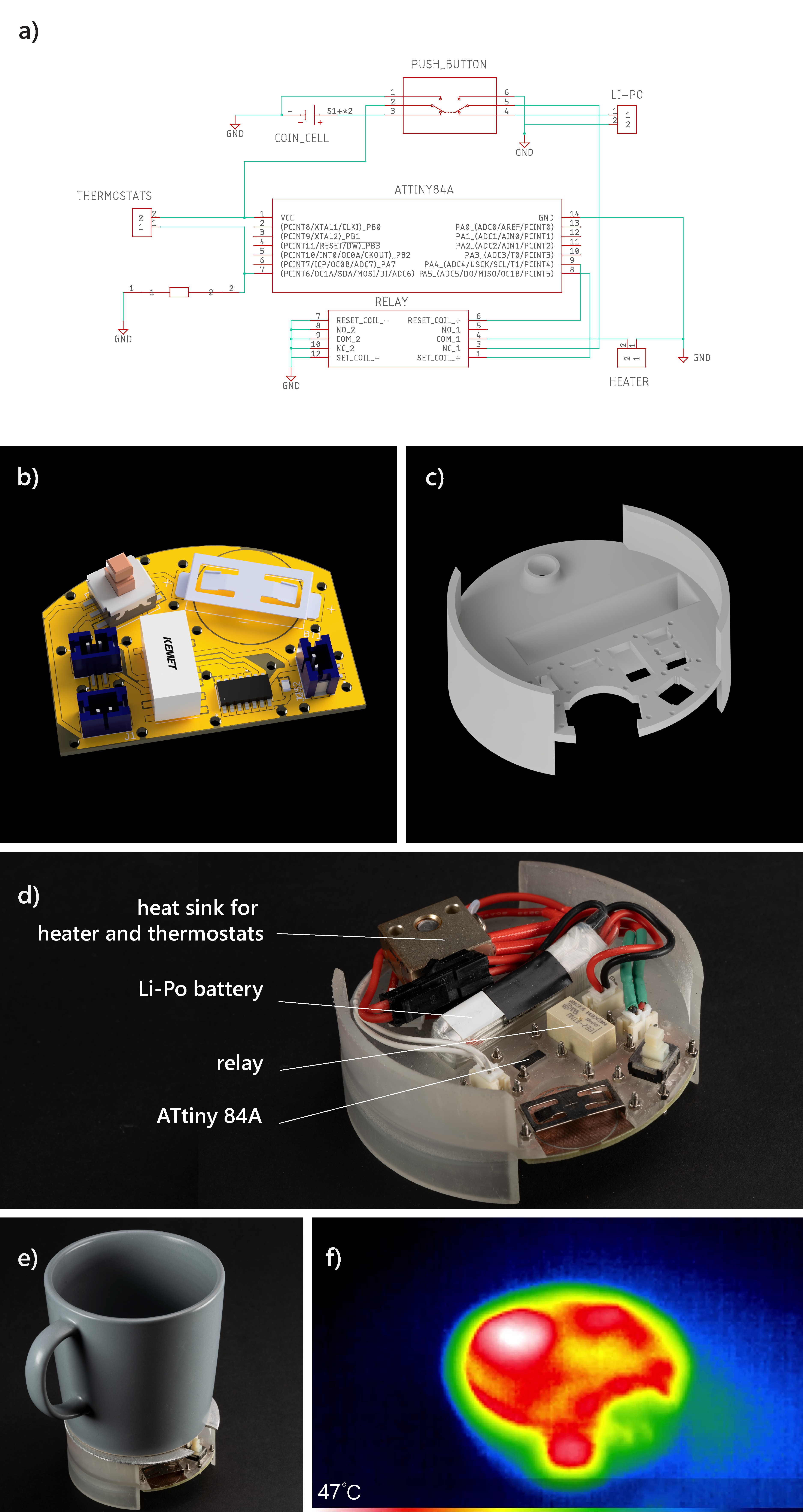}
  \caption{Mug heater made with SolderlessPCB approach: a) the schematic design for the heater, b) a rendering of the heater PCB, c) a rendering of the PCB housing that integrated with the heater's cylinder frame, d) the assembled heater, e) a mug being heated on the heater, f) the thermal imaging of the heater surfaces.}
  \Description{Figure 21: This figure contains six images labeled in 'a', 'b', 'c', 'd', 'e', and 'f' showing a mug heater example and its made process with SolderlessPCB approach. Image 'a' shows the schematic design for the heater. Image 'b' shows a rendering of the heater PCB. Image 'c' shows a rendering of the PCB housing integrated with the heater's cylinder frame. Image 'd' shows an illustration of the assembled heater. Image 'e' shows a mug being heated on the heater. Image 'f' shows a thermal imaging of the heater surfaces with a red heating area.}
  \label{fig:heater}
\end{figure}

\subsubsection{Mug heater}

SolderlessPCB can be used in circuits that deliver higher current beyond those used in typical logic signaling.
Here, we demonstrate a tabletop mug heater designed to maintain liquid temperature.
The mug heater composes a 3D printed mug base plate, an aluminum topping, and a PCB made using SolderlessPCB approach. 
As illustrated in Figures~\ref{fig:heater}a, b, and d, the main component of the PCB is a \SI{12}{\volt}, \SI{60}{\watt} heating element. This element is undervolted and powered by a \SI{7.4}{\volt}, \SI{2500}{\milli\ampere\hour} Li-Po battery with a 35C rating, and is controlled through a relay.

The logic circuit is made with an ATtiny84A microcontroller in an SOIC package and powered by a 2032 coin cell battery. It reads the temperature of the aluminum and controls the relay accordingly. 
The PCB housing design is directly integrated into the cylindrical frame of the mug heater and was manufactured as a single piece, as shown in Figure \ref{fig:heater}c. 
Upon turning on the device, the aluminum surface of the mug heater quickly reaches \SI{40}{\celsius} (as shown in Figure \ref{fig:heater}e and f) within 1 minute and then stabilizes at around \SI{45}{\celsius}. The current draw during heating is around \SI{3.3}{\ampere}.

\subsubsection{Game console}

\begin{figure}[h]
  
  \includegraphics[width=\columnwidth]{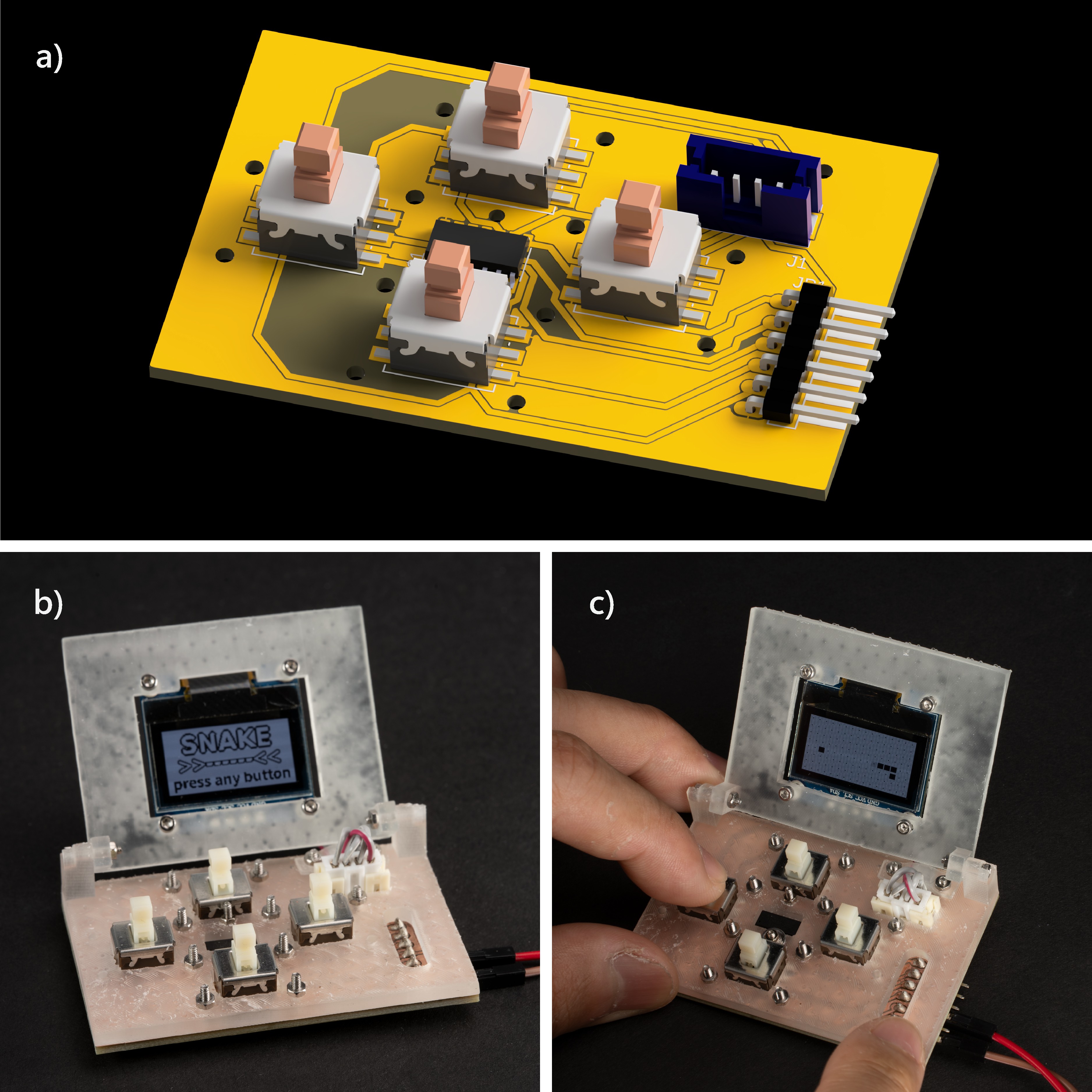}
  \caption{Game console: a) rendering of the game console PCB design, b) the front page of the snake game displayed on the OLED display, c) player gaming with the console.}
  \Description{Figure 22: This figure contains three images labeled in 'a', 'b' and 'c' showing the design of a game console. Image 'a' shows a rendering of the game console PCB design. Image 'b' shows the front page of the snake game displayed on the OLED display when the console is on. Image 'c' shows a player gaming with the console.}
  \label{fig:snake}
\end{figure}

In this example, we utilized the SolderlessPCB approach to create a mini game console, where the display data was transmitted through the I\textsuperscript{2}C protocol.
The PCB comprises an ATtiny84A microcontroller, four non-lock push buttons, and a 4-pin SMD JST connector that interfaces with an OLED display (Figure \ref{fig:snake}a).
The housing of the SolderlessPCB is assembled as part of the game console, with the top housing holding the OLED display screen while the bottom houses the circuits and the buttons (Figure \ref{fig:snake} b and c).
We implemented a snake game operated on the OLED display through I\textsuperscript{2}C, transmitting data at a rate of \SI{100}{\kilo\hertz} with no data loss. The example demonstrates that SolderlessPCB can be used for prototyping circuits that require fast-speed data transmission.

\subsubsection{FTDI unit}
\begin{figure}[h]
  
  \includegraphics[width=\columnwidth]{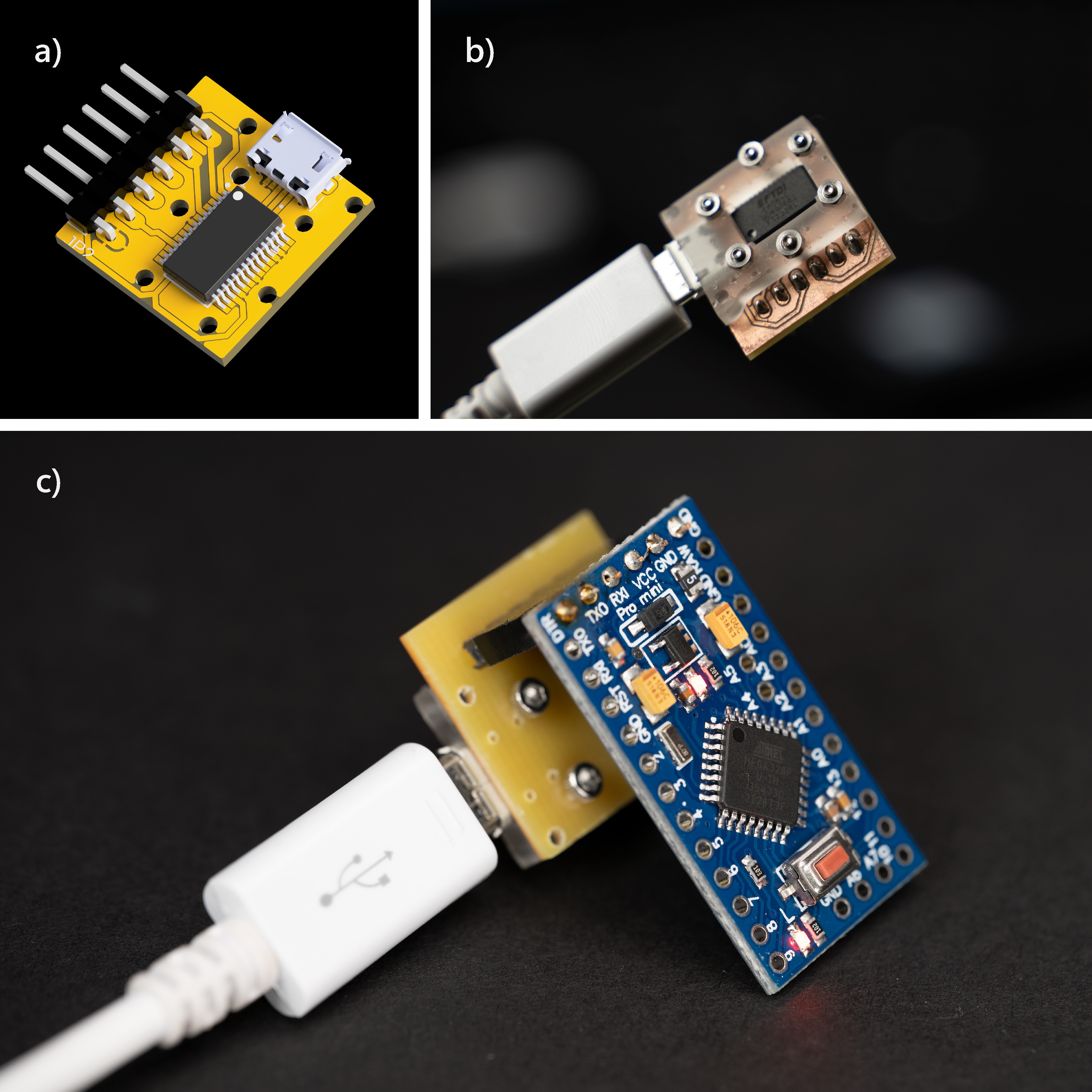}
  \caption{FTDI unit: a) rendering of the PCB design of FTDI unit, b) FTDI unit assembled and connected to USB, c) FTDI unit successfully uploading code to an Arduino Pro Mini.}
  \Description{Figure 23: This figure contains three images labeled in 'a', 'b' and 'c' showing a FTDI unit. Image 'a' shows a rendering of the PCB design of FTDI unit. Image 'b' shows the FTDI unit being assembled and connected to USB. Image 'c' shows FTDI unit successfully uploading code to an Arduino Pro Mini.}
  \label{fig:ftdi}
\end{figure}

In this example, we show that the SolderlessPCB approach can reliably support ICs transmitting data at even higher speeds. 
Future Technology Devices International (FTDI) modules, which convert USB to serial communication, are commonly used for code uploading in miniature Arduino boards lacking on-board UART (universal asynchronous receiver/transmitter) support (Figure \ref{fig:ftdi}a).
Here, we built a custom miniature FTDI adapter using SolderlessPCB, with a 28-pin FTDI232RL IC in an SSOP package and an SMD micro-USB connector. The 3D printed housing secures both the IC and the USB connector with six bolts (Figure \ref{fig:ftdi}b).
The custom FTDI module, made with the SolderlessPCB approach, tolerates repeated plugging and unplugging of the USB cable, and can communicate with the USB at the default rate of \SI{12}{Mbps} and reliably upload code to an Arduino Pro mini (Figure \ref{fig:ftdi}c).

\section{discussion}

\subsection{Housing Waste and Alternatives}
The key idea of SolderlessPCB is to facilitate the easy swapping of SMD components on a PCB prototyping board, so that still-functional electronic components can be salvaged from an old design and reused in the future. While our method promotes the reuse of a wide range of different SMD components, the use of resin-based 3D printed housing requires additional material consumption, which is one limitation of our method.

To mitigate this limitation, we are exploring alternative materials that can be biodegradable and recyclable, thereby reducing the potential for material waste from the housing. Specifically, we explored two additional housing fabrication materials: PLA~\cite{taib2023review}, used with FDM 3D printing, and MDF~\cite{irle2019advanced}, used with CNC milling.

\begin{figure}[h]
  
  \includegraphics[width=\columnwidth]{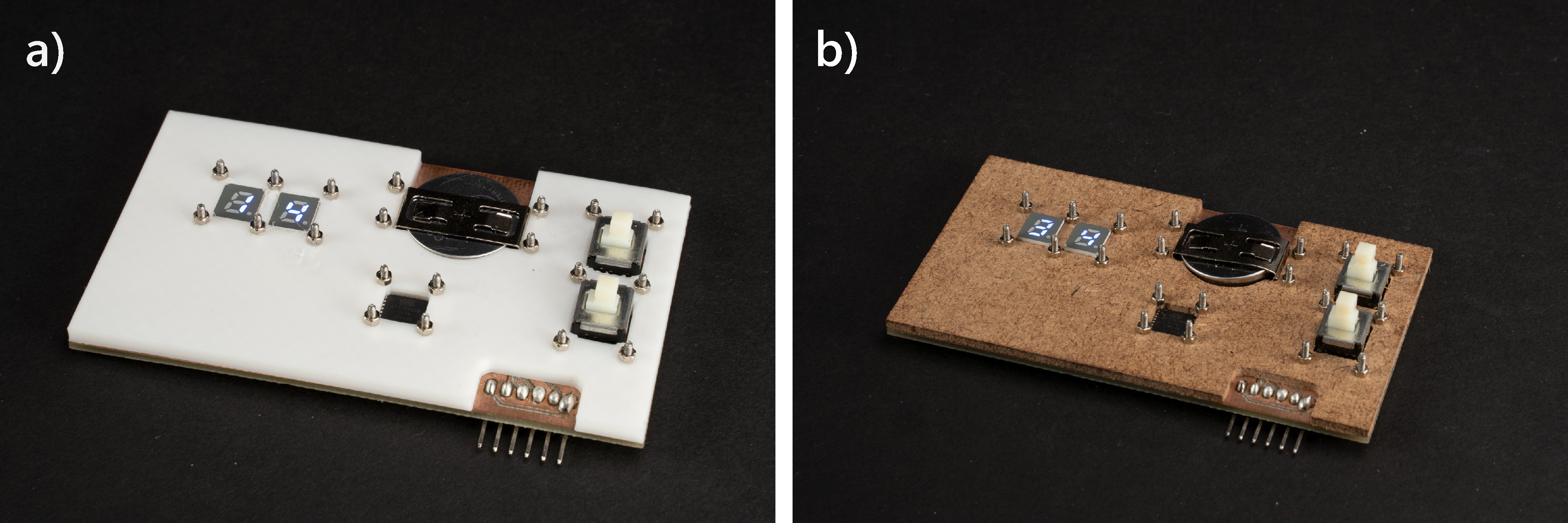}
  \caption{Custom housings in alternative materials: a) PLA housing printed with an FDM 3D printer, b) MDF housing manufactured with a CNC milling machine.}
  \Description{Figure 24: This figure contains two images labeled in 'a' and 'b' showing the housing can be made in in alternative materials. Image 'a' shows a PLA housing printed with an FDM 3D printer. Image 'b' shows a MDF housing manufactured with a CNC milling machine.}
  \label{fig:FDMCNC}
\end{figure}

As one example, we replicated the same housing design for the previous kitchen timer example (Section ~\ref{timer}) using both methods. The kitchen timer PCB contains six SMD electronic components but no small-scale, two-terminal electronic components. The PLA housing was 3D printed with a Bambu Lab X1 3D printer using a 0.2 mm nozzle, while the CNC housing was milled using a Bantam Tools desktop PCB milling machine. We report that both housings worked with the timer PCB baseboard without additional changes to the housing design. This indicates that the SolderlessPCB method could potentially be extended to other materials for hosting ICs and larger SMD components such as buttons and displays (Figure \ref{fig:FDMCNC}). However, it should be noted that we were unable to create the small tabs design for two-terminal components with either FDM 3D printing or CNC milling, as these tabs require a higher precision than either of the machines can achieve.

\subsection{Oxidization}

Throughout the development of SolderlessPCB, we observed a certain level of oxidation on the FR-4 surfaces, which could potentially affect the solderless connection between the oxidized copper and electronic components. To understand how the assembly process influences material oxidation, we compared the oxidation levels of two bare FR-4 boards, one handled by an operator with gloves and the other without. The results show that when there is no direct skin contact, the FR-4 does not exhibit signs of oxidation even after a month of daily handling. In contrast, the FR-4 handled with bare hands begins to oxidize within two weeks. It should be noted that the oxidized prototype remained functional despite signs of oxidation. However, we recommend wearing gloves when directly handling the FR-4 PCBs during the assembly process.

\subsection{Working with Outsourced PCB}

So far, all the PCB baseboards discussed in our paper have been fabricated with FR-4 and machined using a CNC milling machine. To further explore the applicability of the SolderlessPCB approach, we investigated its compatibility with outsourced PCBs. We outsourced PCBs for the kitchen timer, featuring a lead-free HASL (Hot Air Solder Leveling) surface finish. These outsourced PCBs were then encased in three different materials: DLP-printed resin housing, CNC-milled MDF housing, and FDM-printed PLA housing. We report that all housings facilitated reliable connections, as demonstrated in Figure \ref{fig:out}b and c. This suggests that SolderlessPCB can also be utilized with outsourced PCBs, making it applicable to engineers and designers who may not have the capability to fabricate PCBs in-house.

\begin{figure}[h]
  
  \includegraphics[width=\columnwidth]{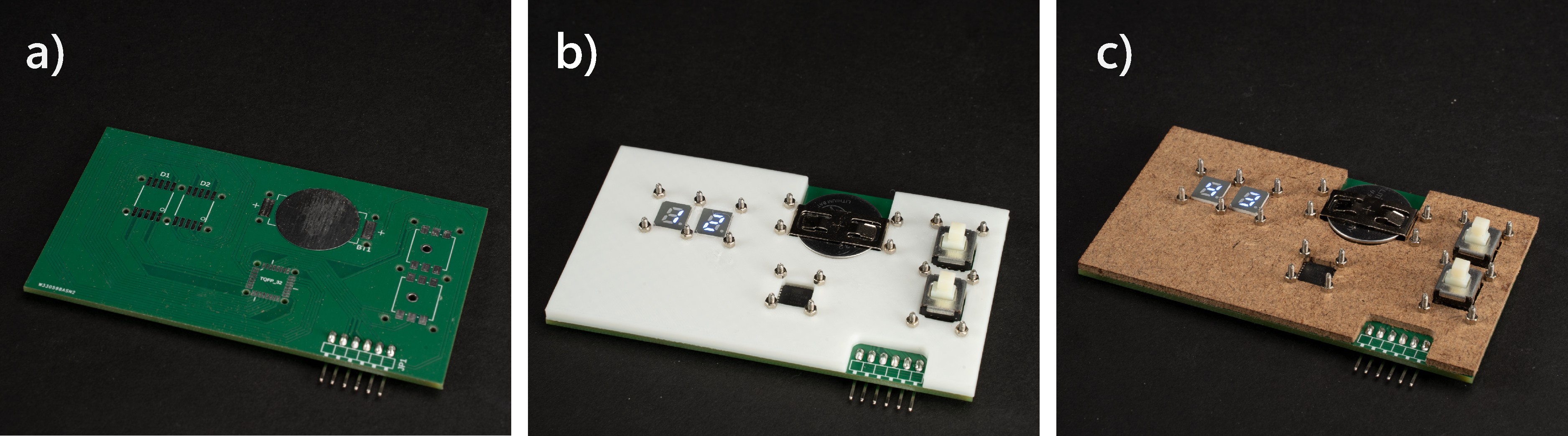}
  \caption{SolderlessPCB working with outsourced PCBs: a) an outsourced PCB with the same layout of the timer example, b) the outsourced PCB working with FDM 3D-printed housing, c) the outsourced PCB working with CNC-milled MDF housing.}
  \Description{Figure 25: This figure contains three images labeled in 'a', 'b' and 'c' showing the SolderlessPCB approach working with outsourced PCBs. Image 'a' shows an outsourced PCB with the same layout of the timer example. Image 'b' shows the outsourced PCB working with FDM 3D-printed housing. Image 'c' shows the outsourced PCB working with CNC-milled MDF housing.}
  \label{fig:out}
\end{figure}

\subsection{Adding Thickness to PCB Design}\label{thickness}
Our method requires an additional housing for a PCB prototype, resulting in an increase in the overall thickness. This is a limitation of our approach when the thickness of the PCB is a critical design constraint.

However, for certain designs, it is possible to integrate the PCB housing as part of the overall form factor, mitigating the effect caused by the housing thickness. For example, in the third iteration of the bristlebot toy (Figure \ref{fig:wt2}c and e), the bot's PCB housing was directly integrated into the toothbrush bases with a single print. Similarly, the mug heater and its PCB housing shown in Figure \ref{fig:heater}c were also integrated as one piece. We expect future research to automate this process, suggesting designs that combine the SolderlessPCB housing and the surrounding 3D-printed cases when necessary.

\subsection{Assembly Effort}\label{assembly effort}
SolderlessPCB eliminates the need for soldering and desoldering of SMD components; instead, the primary assembly and disassembly efforts involve screwing and unscrewing the bolts. 
As anecdotal evidence, we measured the time it took for the first author to assemble and disassemble the scoreboard (Figure \ref{fig:wt1}), which has the highest number of bolts (24) among all the examples presented in the paper.
We report that the assembly process took 5 minutes while disassembly took 3 minutes. 
It should be noted that the effort required for manual (dis)assembly may vary among individuals, and our timing represents just one data point. 
Generalizing the assembly effort requires further investigation.

\begin{figure}[h]
  
  \includegraphics[width=\columnwidth]{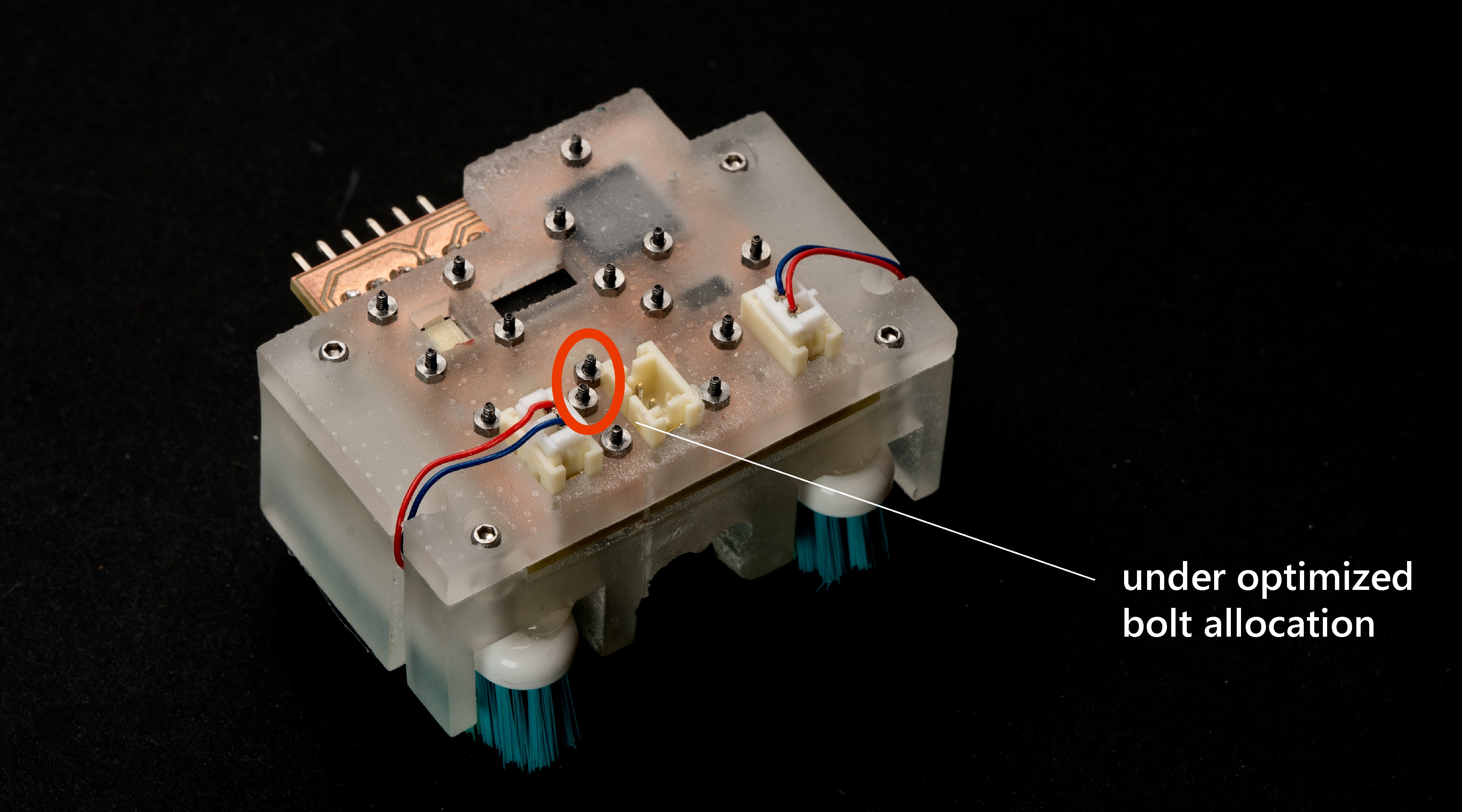}
  \caption{Closely adjacent bolts used in the bristlebot example.}
  \Description{Figure 26: This image shows closely adjacent bolts used in the bristlebot example.}
  \label{fig:out}
\end{figure}

\subsection{Design Automation and Software}
The main focus of this paper is on a new technique that enables the assembly of PCBs without soldering. As mentioned in Section~\ref{workflow}, we have constructed, and open-sourced a custom Fusion 360 component library to ensure SolderlessPCB design compatibility with a variety of existing, off-the-shelf SMD components. This library can be further expanded by the PCB prototyping community.

Looking ahead, we envision opportunities for a more advanced, end-to-end software pipeline that can optimize housing generation and reduce design effort for designers. For example, future software could determine the placement of bolts by holistically considering all SMD components. This approach could potentially lessen the need for bolts compared to our current method (Figure~\ref{fig:out}). As discussed in Section~\ref{thickness}, another feature of future software could be the automatic integration of PCB housings into the overall design form factor when necessary. Finally, we anticipate that future software may assist in counting the number of old SMD components made possible for reuse through SolderlessPCB. This may encourage and promote the practice of reusing electronic components within the community.

\subsection{Retrofitting Current PCB Prototyping Workflow}
As the assembly and disassembly process is the only major difference between SolderlessPCB and conventional PCB prototyping methods, we expect that SolderlessPCB can be integrated into the daily PCB prototyping workflow with minimal effort. To validate this expectation, we plan to conduct a longitudinal study by deploying SolderlessPCB among electronic designers in the near future. We anticipate that the qualitative results from this study will provide insights into how designers can incorporate SolderlessPCB into their daily routines, and shed light on its potential to facilitate the recycling and reuse of electronic components in prototyping practices.

\section{Conclusion}

We have presented SolderlessPCB, a suite of techniques specifically designed to support PCB prototyping without the need for soldering. We detail the design and fabrication guidelines, along with parameter characterizations, to ensure reliable PCB assemblies. We showcase two sets of scenarios and three additional examples, which demonstrate both the salvage and reuse of SMD components and the diverse types of circuits that can be realized using SolderlessPCB. We conclude with discussions on the limitations of our approach, potential methods to mitigate these limitations, and future research directions.

\begin{acks}
We thank the Anlage Research Group at the University of Maryland for sharing access to their measurement equipment, and Jingnan Cai for providing the necessary training. We also thank the Jagdeep Singh Family Makerspace for providing access to tools during the development and documentation processes of this project. ChatGPT is used in this work solely to correct spelling or grammar errors.
\end{acks}

\bibliographystyle{ACM-Reference-Format}
\bibliography{sample-base}

\end{document}